\documentclass[usenatbib]{mn2e} 
\usepackage{graphicx,fixltx2e,amsmath,amssymb,amsfonts,latexsym,wasysym}

\def\rpd{\hbox{rad\,d$^{-1}$}}

\def\chisqr{\hbox{$\chi^2_{\rm r}$}}
\def\msun{\hbox{${\rm M}_{\odot}$}}
\def\mjup{\hbox{${\rm M}_{\jupiter}$}}
\def\mspy{\hbox{${\rm M}_{\odot}$\,yr$^{-1}$}}
\def\rsun{\hbox{${\rm R}_{\odot}$}}
\def\lsun{\hbox{${\rm L}_{\odot}$}}
\def\rcor{\hbox{$r_{\rm cor}$}}
\def\rmag{\hbox{$r_{\rm mag}$}}
\def\mstar{\hbox{$M_{\star}$}}
\def\rstar{\hbox{$R_{\star}$}}

\def\teff{\hbox{$T_{\rm eff}$}}
\def\logg{\hbox{$\log g$}}
\def\sn{\hbox{S/N}}
\def\vrad{\hbox{$v_{\rm rad}$}}

\def\kms{\hbox{km\,s$^{-1}$}}
\def\vsini{\hbox{$v \sin i$}}

\def\ptt{\hbox{$10^{-4} I_{\rm c}$}}
\def\mV{\hbox{$m_{\rm V}$}}

\def\degr{\hbox{$^\circ$}}

\def\omeq{\hbox{$\Omega_{\rm eq}$}}
\def\dom{\hbox{$d\Omega$}}

\newcommand{\caii}{\hbox{Ca$\;${\sc ii}}}
\newcommand{\fei}{\hbox{Fe$\;${\sc i}}}

\newcommand{\hei}{\hbox{He$\;${\sc i}}}

\begin{document}

\title[Dynamo \& accretion processes of V2129~Oph]{Non-stationary dynamo and magnetospheric 
accretion processes of the classical T~Tauri star V2129~Oph}  
\makeatletter

\def\newauthor{%
  \end{author@tabular}\par
  \begin{author@tabular}[t]{@{}l@{}}}
\makeatother
 
\author[J.-F.~Donati et al.]
{\vspace{1.7mm}
J.-F.~Donati$^1$\thanks{E-mail: 
donati@ast.obs-mip.fr }, 
J.~Bouvier$^2$, F.M.~Walter$^3$, S.G.~Gregory$^{4}$, M.B.~Skelly$^1$, \\ 
\vspace{1.7mm}
{\hspace{-1.5mm}\LARGE\rm 
G.A.J.~Hussain$^5$, E.~Flaccomio$^6$, C.~Argiroffi$^7$, K.N.~Grankin$^8$, M.M.~Jardine$^9$,} \\ 
\vspace{1.7mm}
{\hspace{-1.5mm}\LARGE\rm
F.~M\'enard$^2$, C.~Dougados$^2$, M.M.~Romanova$^{10}$ \& the MaPP collaboration} \\ 
$^1$ LATT--UMR 5572, CNRS \& Univ.\ de Toulouse, 14 Av.\ E.~Belin, F--31400 Toulouse, France \\
$^2$ LAOG--UMR 5571, CNRS \& Univ.\ J.~Fourier, 414 rue de la Piscine, F--38041 Grenoble, France \\ 
$^3$ Department of Physics and Astronomy, Stony Brook University, Stony Brook NY 11794--3800, USA \\ 
$^4$ School of Physics, Univ.\ of Exeter, Stocker Road, Exeter EX4~4QL, UK \\ 
$^5$ ESO, Karl-Schwarzschild-Str.\ 2, D-85748 Garching, Germany \\ 
$^6$ INAF, Osservatorio Astronomico di Palermo, Piazza del Parlamento 1, 90134 Palermo, Italy \\ 
$^7$ DSFA, Univ.\ di Palermo, Piazza del Parlamento 1, 90134 Palermo, Italy \\ 
$^8$ Crimean Astrophysical Observatory, Nauchny, Crimea 334413, Ukraine \\ 
$^9$ School of Physics and Astronomy, Univ.\ of St~Andrews, St~Andrews, Scotland KY16 9SS, UK \\
$^{10}$ Department of Astronomy, Cornell University, Ithaca, NY 14853-6801, USA 
}

\date{2010 November, MNRAS in press}
\maketitle
 
\begin{abstract}  

We report here the first results of a multi-wavelength campaign focussing 
on magnetospheric accretion processes of the classical T~Tauri star (cTTS) V2129~Oph.  
In this paper, we present spectropolarimetric observations collected in 2009 July with ESPaDOnS 
at the Canada-France-Hawaii Telescope (CFHT) and contemporaneous photometry secured with the SMARTS 
facility.  Circularly polarised Zeeman signatures are clearly detected, both in photospheric 
absorption and accretion-powered emission lines, from time-series of which we reconstruct 
new maps of the magnetic field, photospheric brightness and accretion-powered emission at 
the surface of V2129~Oph using our newest tomographic imaging tool -- to be compared with 
those derived from our old 2005 June data set, reanalyzed in the exact same way.  

We find that in 2009 July, V2129~Oph hosts octupolar and dipolar field components of about 
2.1 and 0.9~kG respectively, both tilted by about 20\degr\ with respect to the rotation 
axis;  we conclude that the large-scale magnetic topology changed significantly since 
2005 June (when the octupole and dipole components were about 1.5 and 3 times weaker 
respectively), demonstrating that the field of V2129~Oph is generated by a non-stationary 
dynamo.  We also show that V2129~Oph features a dark photospheric spot and a localised area 
of accretion-powered emission, both close to the main surface magnetic region 
(hosting fields of up to about 4~kG in 2009 July).  We finally obtain that the surface shear of V2129~Oph 
is about half as strong as solar.  

From the fluxes of accretion-powered emission lines, we estimate that the observed average logarithmic 
accretion rate (in \mspy) at the surface of V2129~Oph is $-9.2\pm0.3$ at both epochs, peaking 
at $-9.0$ at magnetic maximum.  It implies in particular that the radius at which the 
magnetic field of V2129~Oph truncates the inner accretion disc is 0.93$\times$ and 0.50$\times$ 
the corotation radius (where the Keplerian period equals the stellar rotation period) in 2009 
July and 2005 June respectively.  
\end{abstract}


\section{Introduction} 
\label{sec:int}

Magnetic fields are thought to play an important role during the formation of stars 
and their planetary systems.  In particular, they presumably oppose the collapse of 
the giant molecular clouds from which stars and planets form, strongly inhibit its 
fragmentation into multiple globules, efficiently evacuate most of the accreting mass 
(e.g., through conical outflows and collimated jets) and rapidly dissipate the 
initial reservoir of angular momentum \citep[e.g.,][]{Andre08};  they also likely 
affect the formation and migration of planets in the accretion discs surrounding 
newly-born stars.  

The impact of magnetic fields on the formation of low-mass stars (i.e., with 
$\mstar\leq3$~\msun) that are still surrounded with a gaseous accretion disc --  
the so-called classical T~Tauri stars or cTTSs -- has been the subject of intense 
scrutiny and vivid debates in the last two decades.  With their intense 
large-scale fields, cTTSs are putatively capable of disrupting the inner regions 
of their accretion discs and of braking their rotation rather drastically, 
although the exact physical process through which they can reach this result 
is still controversial \citep[e.g.,][for a review]{Bouvier07}.  

Following several papers reporting the presence of strong fields  on cTTSs 
\citep[e.g.,][]{Johns07}, the first spectropolarimetric studies based on time-series of 
Zeeman signatures from accreting and non-accreting regions at the surfaces of cTTSs 
revealed that low-mass protostars indeed possess large-scale fields, but that the 
intensity and topologies of their fields strongly depend on the protostellar mass 
in a way very reminiscent to those of low-mass main-sequence stars \citep{Donati07, 
Donati08, Donati09, Donati10}.  These initial results however call for a confirmation 
through surveys on a statistically significant sample of cTTSs, allowing in particular to 
study how the properties of their large-scale magnetic fields depend on stellar parameters 
such as mass, age, rotation and accretion rates.  Ultimately, such investigations can guide 
us towards more realistic models of forming Suns, taking into account the effect of magnetic 
fields in particular.  

Magnetic Protostars and Planets (MaPP) is an international project focussing 
specifically on this issue.  It has been granted 690~hr of observing time over 9 
consecutive semesters (2008b to 2012b) with the ESPaDOnS spectropolarimeter on 
the 3.8-m Canada-France-Hawaii Telescope (CFHT) to survey 15~cTTSs and 3 protostellar 
accretion discs of FU~Ori type (FUOrs);  it also regularly benefits from 
contemporaneous observations with the NARVAL spectropolarimeter on the 2-m T\'elescope 
Bernard Lyot (TBL) as well as photometric observations from Crimea, Uzbekistan and 
Armenia.  Additional multiwavelength observations from space 
and/or from the ground are also organised in conjunction with MaPP 
campaigns on a few specific targets, providing deeper insights into the physical processes 
under scrutiny (and in particular magnetospheric accretion).  

Following a first MaPP paper concentrating on the prototypical low-mass cTTS AA~Tau 
and suggesting that its large-scale magnetic field is strong enough to expel outwards 
(e.g., through a conical wind) most of the accreted disc material (the propeller 
regime) and thus to spin-down the protostar efficiently \citep{Donati10b},  
we present a new study revisiting the properties of the large-scale 
magnetic field of V2129~Oph \citep[after the intial investigation of][]{Donati07};  
in addition to being based on a much better and longer spectropolarimetric data 
set, this study also benefits from additional contemporaneous multiwavelength 
observations from both space (at X-ray frequencies with Chandra/HETGS) and from the 
ground (high-resolution spectroscopy from HARPS@ESO, low resolution 
spectroscopy and photometry from the SMARTS facility).  A brief overview of the goals of the
multiwavelength observing program is given in \citet{Gregory09}.  

The present paper concentrates mostly on the spectropolarimetric data, while the other 
related data sets (and in particular the Chandra/HETGS spectra, Argiroffi et al, in 
preparation) will be described and analysed in 
forthcoming companion papers.  After briefly recalling the stellar parameters of 
V2129~Oph, (see Sec.~\ref{sec:v2129}), we describe the observed variations of photospheric 
lines and accretion proxies (Sec.~\ref{sec:var}) and detail their subsequent modelling 
with our magnetic imaging code (Sec.~\ref{sec:mod});  we also include a complete remodelling 
of our older data to ease the comparison of the magnetic images at both epochs.  
We finally discuss the implications of these new results for our understanding of 
magnetospheric accretion processes in cTTSs (Sec.~\ref{sec:dis}).

\section{V2129~Oph = SR~9 = ROX~29}
\label{sec:v2129}

V2129~Oph is the brightest cTTS in the $\rho$~Oph star formation cloud, with limited visual 
extinction ($\simeq$0.6) and a photospheric temperature of $\simeq$4500~K \citep[spectral 
type K5,][]{Donati07}.  
It exhibits strong H$\alpha$ emission typical to cTTSs \citep[e.g.,][]{Bouvier92} and 
clear infrared excesses and 10~$\mu$m silicate emission line fluxes typical of dusty 
accretion discs \citep[e.g.,][]{Furlan09} suggesting a disc mass of $\simeq$1~\mjup\ 
\citep{Cieza10}.  

The distance to $\rho$~Oph has recently been redetermined with improved accuracy using 
VLBA, and estimated to $120\pm5$~pc \citep{Loinard08}.  From the observed maximum 
visual brightness of V2129~Oph \citep[$\mV\simeq11.2$,][]{Grankin08}, the estimated 
unspotted visual magnitude \citep[$11.0\pm0.2$,][]{Donati07} and the visual bolometric 
correction adequate to V2129~Oph \citep[$\simeq$$-1.2$,][]{Bessell98}, one can derive a 
logarithmic luminosity (with respect to the Sun) of $0.15\pm0.1$ and thus a radius of 
$\rstar=2.0\pm0.3$~\rsun.  This is slightly smaller than that derived in the previous study, 
directly reflecting the shorter distance to $\rho$~Oph used here.  

The rotation period of V2129~Oph estimated from regular photometric variations is found 
to be 6.53~d in average, varying from about 6.35 to 6.60~d depending on the observing season 
\citep{Grankin08};  when compared to other similar stars of the same sample (e.g., V410~Tau, 
exhibiting no such behaviour), we conclude that this period fluctuation is real.  It 
suggests that 
differential rotation is shearing the spotted photosphere, producing the observed period 
fluctuations as starspots migrate in latitude on long timescales -- a suggestion that 
we independently confirm in our study (see Sec.~\ref{sec:mod}).  
From the rotational broadening of line profiles, the line-of-sight projected equatorial 
rotation velocity of V2129~Oph is found to be $\vsini=14.5\pm0.3$~\kms\ \citep{Donati07}, 
where $i$ is the inclination of the rotation axis to the line of sight.  
It implies that $\rstar \sin i \simeq 1.8$~\rsun, assuming that the rotation period at 
the equator is close to the lowest observed periods (i.e., that the photospheric shear 
of V2129~Oph is Sun-like, with an equator rotating faster than the poles, as in most other 
low-mass stars showing differential rotation).  

Coupling with the above radius estimate, it suggests that $i\simeq65\degr$, with potential 
values ranging from 50\degr\ to 90\degr.  Tomographic imaging (including this study) suggests 
a smaller inclination angle of $i\simeq45\pm15\degr$ \citep[][]{Donati07}.  
We therefore use an intermediate value, $i\simeq60\degr$, slightly larger than in the 
previous study as a result of the shorter distance assumed for $\rho$~Oph.  
The corresponding value of the stellar radius is $\rstar\simeq2.1$~\rsun\ which, when coupled 
to the temperature, yields a mass of $\mstar\simeq1.35$~\msun\ and an age of about $2-3$~Myr 
\citep{Siess00};  models suggest in addition that V2129~Oph is no longer fully convective 
and hosts a core whose fractional mass and radius are about one third those of the star.  
In this context, the corotation radius (at which the Keplerian period is equal to
the rotation period of the star) is equal to $\rcor\simeq7.7$~\rstar\ or 0.076~AU.

Mass accretion on V2129~Oph is only moderate.  From the equivalent widths and the 
corresponding line fluxes of the emission lines usually considered as accretion proxies 
(and in particular the \hei\ $D_3$ line and the \caii\ infrared triplet, IRT) and using 
the published empirical correlations between lines and accretion fluxes \citep{Fang09}, 
we can derive an estimate of the average logarithmic mass accretion rate (in \mspy) at the 
surface of V2129~Oph, that we find to be equal to $-9.2\pm0.3$\footnote{Note that this new 
estimate is an order of magnitude smaller than that mentioned in our previous study 
\citep{Donati07}, at the time simply derived by averaging measurements obtained with different 
methods and thus likely far less accurate;  in particular, our new mass accretion rate, computed 
in exactly the same way as those of V2247~Oph \citep{Donati10} and AA~Tau \citep{Donati10b}, 
is at least much more consistent with our latter MaPP studies.}.  
Our data also suggest that the mass accretion rate of V2129~Oph is more 
stable on the long-term (see Sec.~\ref{sec:var}) than that of cTTSs like 
AA~Tau \citep[showing dominant intrinsic variability on short timescales,][]{Donati10b};  
this preliminary conclusion however needs to be confirmed with longer data sets, to check 
in particular that we did not catch V2129~Oph in an unusually quiet state of accretion.  
Optical veiling, i.e., the apparent weakening of the photospheric spectrum 
(presumably caused by accretion) is apparently weak for V2129~Oph and does not exceed a few \%.

\section{Observations}
\label{sec:obs}

Spectropolarimetric observations of V2129~Oph were collected from 2009 July 01 to July 14 
using ESPaDOnS on the CFHT.  ESPaDOnS collects stellar spectra spanning the whole optical domain 
(from 370 to 1,000~nm) at a resolving power of 65,000 (i.e., 4.6~\kms) and with a spectral 
sampling of 2.6~\kms, in either circular or linear polarisation \citep{Donati03}.  
A total of 23 circular polarisation spectra were collected, at a rate of 2 spectra per night 
during the first 10~nights.  
All polarisation spectra consist of 4 individual subexposures lasting 
each 690~s and taken in different polarimeter configurations to allow the removal of 
all spurious polarisation signatures at first order.

All raw frames are processed with {\sc Libre~ESpRIT}, a fully automatic reduction
package/pipeline available at CFHT.  It automatically performs optimal 
extraction of ESPaDOnS unpolarized (Stokes $I$) and circularly polarized (Stokes $V$) 
spectra grossly following the procedure described in \citet{Donati97b}.
The velocity step corresponding to CCD pixels is about 2.6~\kms;  however, thanks
to the fact that the spectrograph slit is tilted with respect to the CCD lines,
spectra corresponding to different CCD columns across each order feature a
different pixel sampling.  {\sc Libre~ESpRIT} uses this opportunity to carry out
optimal extraction of each spectrum on a sampling grid denser than the original
CCD sampling, with a spectral velocity step set to about 0.7 CCD pixel
(i.e.\ 1.8~\kms).
All spectra are automatically corrected of spectral shifts resulting from
instrumental effects (e.g., mechanical flexures, temperature or pressure variations) 
using telluric lines as a reference.  Though not perfect, this procedure provides 
spectra with a relative radial velocity (RV) precision of better than 0.030~\kms\
\citep[e.g.,][]{Donati08b}.

\begin{table}
\caption[]{Journal of observations collected in 2009 July.   
Columns $1-4$ respectively list the UT date, the heliocentric Julian date and 
UT time (both at mid-exposure), and the peak signal to noise ratio (per 2.6~\kms\ 
velocity bin) of each observation (i.e., each sequence of $4\times690$~s subexposures).  
Column 5 lists the rms noise level (relative to the unpolarized continuum level 
$I_{\rm c}$ and per 1.8~\kms\ velocity bin) in the circular polarization profile 
produced by Least-Squares Deconvolution (LSD), while column~6 indicates the 
rotational cycle associated with each exposure (using the ephemeris given by 
Eq.~\ref{eq:eph}).  }   
\begin{tabular}{cccccc}
\hline
Date & HJD          & UT      &  \sn\  & $\sigma_{\rm LSD}$ & Cycle \\
     & (2,455,000+) & (h:m:s) &      &   (\ptt)  & (225+) \\
\hline
Jul 01 & 13.76862 & 06:21:07 & 180 & 2.5 & 0.692 \\
Jul 01 & 13.90207 & 09:33:18 & 180 & 2.8 & 0.712 \\
Jul 02 & 14.77090 & 06:24:30 & 200 & 2.2 & 0.845 \\
Jul 02 & 14.90375 & 09:35:50 & 200 & 2.2 & 0.866 \\
Jul 03 & 15.76152 & 06:11:05 & 190 & 2.3 & 0.997 \\
Jul 03 & 15.89192 & 09:18:52 & 210 & 1.9 & 1.017 \\
Jul 04 & 16.75643 & 06:03:51 & 200 & 2.1 & 1.150 \\
Jul 04 & 16.88363 & 09:07:02 & 210 & 1.9 & 1.169 \\
Jul 05 & 17.75614 & 06:03:32 & 190 & 2.3 & 1.303 \\
Jul 05 & 17.88444 & 09:08:18 & 210 & 2.0 & 1.322 \\
Jul 06 & 18.75731 & 06:05:19 & 100 & 4.4 & 1.456 \\
Jul 06 & 18.88404 & 09:07:49 & 190 & 2.2 & 1.475 \\
Jul 07 & 19.75701 & 06:04:59 & 120 & 4.2 & 1.609 \\
Jul 07 & 19.88577 & 09:10:24 & 160 & 3.1 & 1.629 \\
Jul 08 & 20.75692 & 06:04:57 & 170 & 2.6 & 1.762 \\
Jul 08 & 20.88477 & 09:09:04 & 140 & 3.4 & 1.782 \\
Jul 09 & 21.75541 & 06:02:52 & 150 & 3.0 & 1.915 \\
Jul 09 & 21.88237 & 09:05:42 & 160 & 2.7 & 1.935 \\
Jul 10 & 22.75503 & 06:02:26 & 170 & 2.4 & 2.068 \\
Jul 10 & 22.88263 & 09:06:11 & 180 & 2.3 & 2.088 \\
Jul 12 & 24.78050 & 06:39:19 & 150 & 3.0 & 2.378 \\
Jul 13 & 25.84672 & 08:14:48 & 180 & 2.7 & 2.542 \\
Jul 14 & 26.84682 & 08:15:03 & 150 & 3.0 & 2.695 \\
\hline
\end{tabular}
\label{tab:logesp}
\end{table}

The peak signal-to-noise ratios (\sn, per 2.6~\kms\ velocity bin) achieved on the
collected spectra (i.e., the sequence of 4 subexposures) range between 100 and
210 depending on weather/seeing conditions, with a median of 180.  
Rotational cycles $E$ are computed from heliocentric Julian dates 
according to the ephemeris of \citet{Donati07}:  
\begin{equation}
\mbox{HJD} = 2453540.0 + 6.53 E 
\label{eq:eph}
\end{equation}
The full journal of observations is presented in Table~\ref{tab:logesp}.  
Observations collected on July 04 (rotation cycles 1.150 and 1.169) were 
recorded with the (full) moon at an angular distance of only about 2\degr\ 
from V2129~Oph.  As a result, the corresponding unpolarized spectra show 
significant contamination and were discarded from the study;  
circularly polarised spectra are however unaffected and were kept in the 
analysis (see Sec.~\ref{sec:var} and Fig.~\ref{fig:lsd}).  

Least-Squares Deconvolution \citep[LSD,][]{Donati97b} was applied to all
observations.   The line list we employed for LSD is computed from an {\sc
Atlas9} LTE model atmosphere \citep{Kurucz93} and corresponds to a K5IV 
spectral type ($\teff=4,500$~K and  $\logg=3.5$) appropriate for V2129~Oph.
Only moderate to strong atomic spectral lines (with line-to-continuum core 
depressions larger than 40\% prior to all non-thermal broadening) are included 
in this list;  spectral regions with strong lines mostly formed outside the 
photosphere (e.g., Balmer, He, \caii\ H, K and IRT lines) and/or heavily crowded 
with telluric lines were discarded.  
Altogether, about 8,500 spectral features (with about 40\% from \fei) are used 
in this process.  
Expressed in units of the unpolarized continuum level $I_{\rm c}$, the average 
noise levels of the resulting Stokes $V$ LSD signatures are ranging from 1.9 to 
4.4$\times10^{-4}$ per 1.8~\kms\ velocity bin.  

Almost simultaneous photometry was collected with the SMARTS facility in BVRIJHK 
bands from rotation cycles 224.3 to 226.9, with a total of 10 to 13 data points 
depending on the band.  These data will be 
analysed in detail in a forthcoming companion paper;  we only use here the V band 
measurements as a guideline for checking the consistency of our modelling.

\section{Variations of photospheric lines and accretion proxies}
\label{sec:var}

As in our previous run \citep{Donati07}, Zeeman signatures are detected at all times in 
Stokes $V$ LSD profiles, with peak-to-peak amplitudes varying from 0.2 to 1.0\% (see 
Fig.~\ref{fig:lsd} top right panel).  Temporal variations are dominant, the average 
Stokes $V$ signature being most of the time smaller than the individual ones;  moreover, 
Zeeman signatures are complex and feature several reversals throughout the line profile, 
suggesting that the parent field topology is not simple.  
As a result, the line-of-sight projected component of the field averaged over 
the visible stellar hemisphere and weighted by brightness inhomogeneities 
\citep[called longitudinal field and estimated from the first 
moment of the Stokes $V$ profile, e.g.,][]{Donati97b}, varying from $-360$ to $+90$~G, is 
showing clear non-sinusoidal fluctuations (see Fig.~\ref{fig:var} lower left panel) 
similar to (and about 50\% larger than) those observed in 2005 June.  

LSD Stokes $I$ profiles of photospheric lines, centred at an average RV of about 
$-7$~\kms, exhibit temporal shape variations, and at 
times clear asymmetries, (e.g., at rotation cycles 1.303 and 2.542) very reminiscent to those 
caused by cool spots at the surface of active stars.  This is again consistent with the 
conclusions reached from our first set of observations, where line profiles distortions 
(smaller than but similar to those reported here) and their evolution with time were 
successfully interpreted in terms of cool spots at the surface of V2129~Oph \citep{Donati07}.  
The observed photometric variations (see Fig.~\ref{fig:var} upper panel) and in particular 
the corresponding colors (not shown here, to be published in a forthcoming companion paper) 
also strongly argue that cool spots are present at the 
surface of the star, with maximum spottedness occurring around phase 0.55, i.e., very close 
to the phase of strongest photospheric longitudinal field.  

We also find that Stokes $I$ LSD profiles show equivalent widths nearly equal to those in 
the previous data set and roughly constant with time (within $\simeq2$\%), suggesting that 
veiling is small.  Assuming that the LSD Stokes $I$ profiles with maximum equivalent width 
correspond to an unveiled spectrum, we derive that veiling is always smaller than 5\% 
(see Fig.~\ref{fig:var}), except on July~05 where it reaches about 7\% as a likely result 
of a spectral pollution from the nearby full Moon.  Comparison with spectral lines of 
non-accreting TTSs of similar spectral types (e.g., V410~Tau) confirms this conclusion.  

We find in particular that LSD Stokes $I$ and $V$ profiles repeat well from one 
rotation cycle to the next, suggesting that rotational modulation clearly dominates over 
intrinsic variability.  This is particularly clear on the longitudinal field curve 
(Fig.~\ref{fig:var} lower left panel) that repeats almost identically over the 
2 successive rotation cycles.  In particular, this behaviour is radically different 
from that of AA~Tau where intrinsic variability strongly dominates the photospheric 
longitudinal field curve \citep{Donati10b}.  

\begin{figure*}
\vspace{-4mm}
\center{\hbox{\includegraphics[scale=0.6,angle=-90]{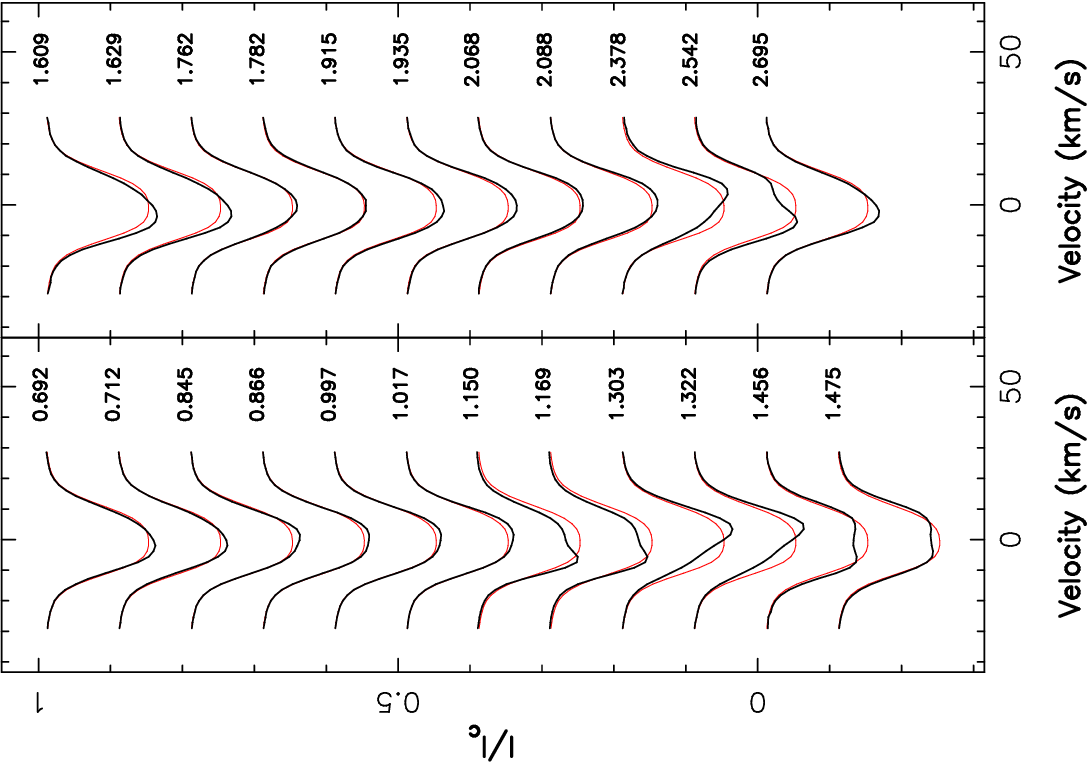}\hspace{4mm}
\includegraphics[scale=0.6,angle=-90]{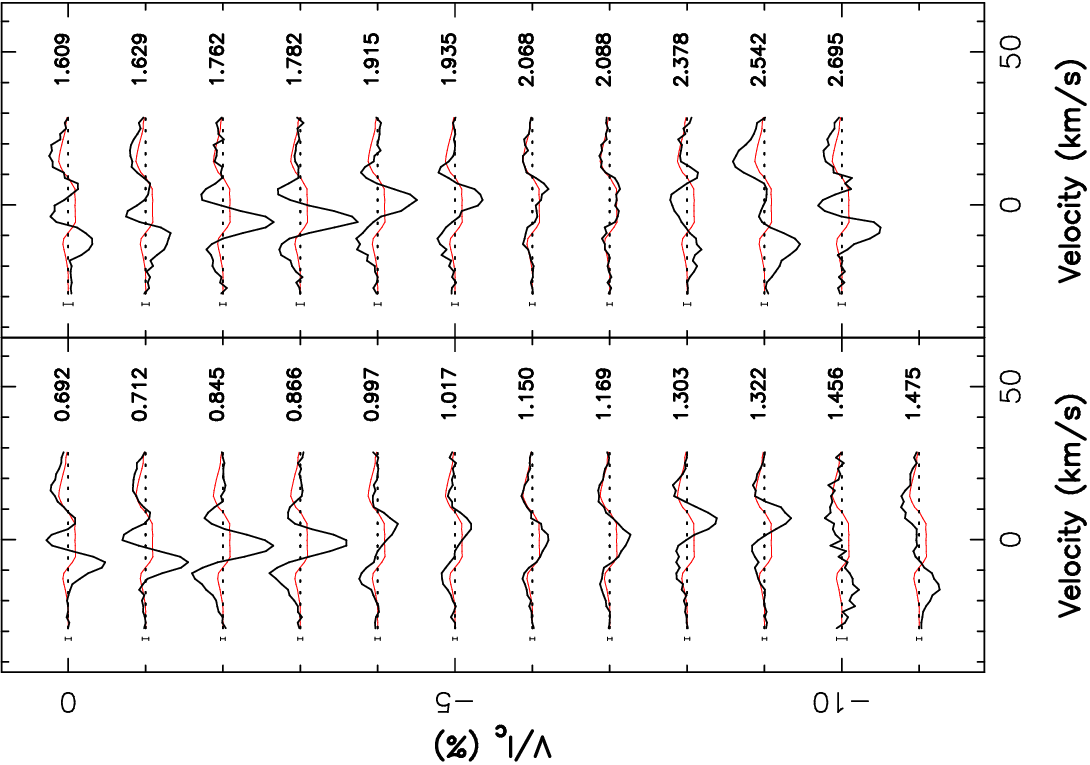}}} 
\vspace{5mm}
\center{\hbox{\includegraphics[scale=0.6,angle=-90]{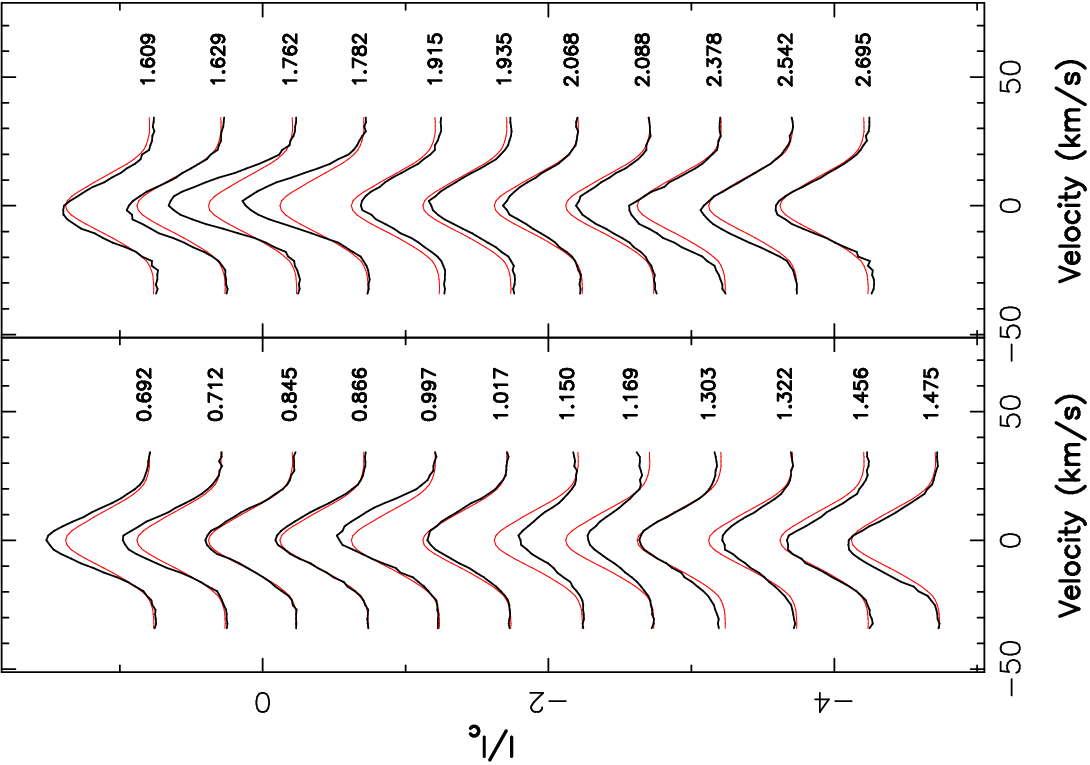}\hspace{4mm}
\includegraphics[scale=0.6,angle=-90]{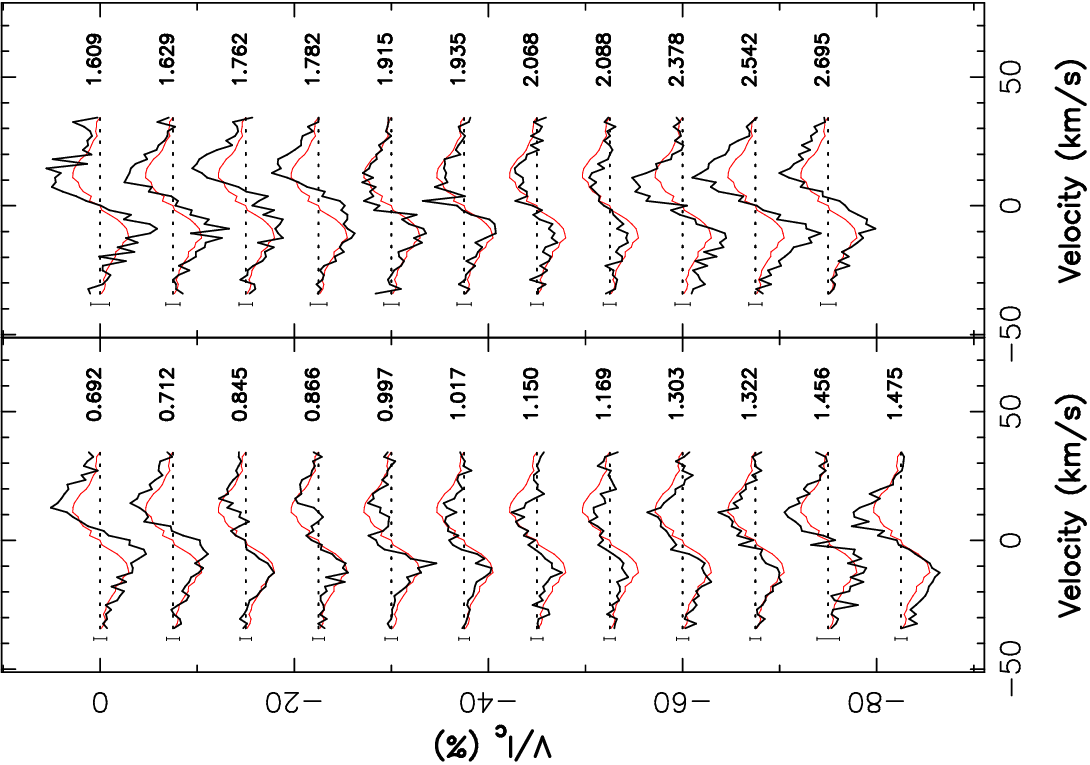}}} 
\caption[]{Variation of the Stokes $I$ (left) and Stokes $V$ (right) LSD profiles 
of the photospheric lines (top) and \caii\ IRT lines (bottom) of V2129~Oph in 2009 July.  
The Stokes $I$ profiles of \caii\ emission is shown before subtracting the underlying (much
wider) photospheric absorption, hence the reduced flux in the far wings (with respect to the
unit continuum).
To emphasize variability, the average profile over the run is shown in red.  
Rotation cycles (as listed in Table~1) and 3$\sigma$ error bars (for Stokes $V$ data 
only) are also included next to each profile.  Unpolarized profiles recorded on 
rotation cycles 1.150 and 1.169 are significantly affected by the moon and were 
discarded from the following study. } 
\label{fig:lsd}
\end{figure*}

\begin{figure*}
\center{\hbox{\includegraphics[scale=1.0,angle=-90]{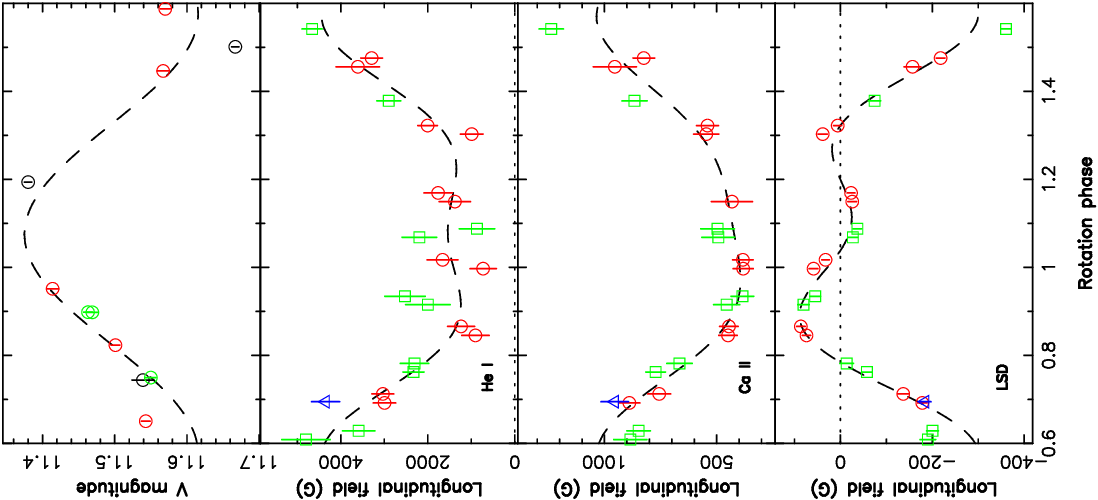}\hspace{4mm}
\includegraphics[scale=1.0,angle=-90]{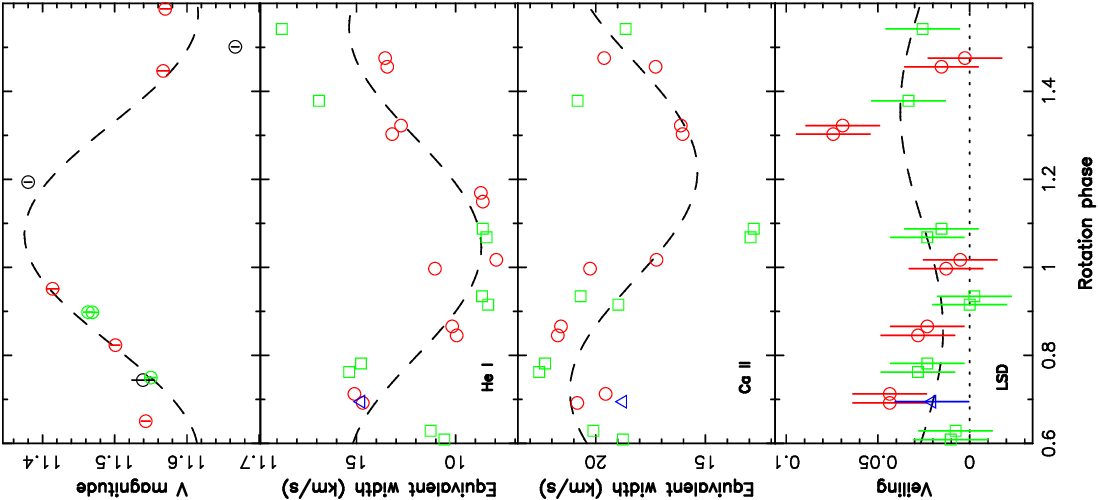}}} 
\caption[]{Temporal variations of the integrated brightness (top panels), \hei\ D3 (second 
panels), \caii\ IRT (third panels) and photospheric lines (bottom panels) of V2129~Oph in 2009 
July.  
Longitudinal field variations are shown on the left side (3 lower panels) while equivalent 
width and veiling variations are shown on the right side.  Data 
collected within cycles $0.6-1.6$, $1.6-2.6$ and $2.6-3.6$ are respectively plotted 
as red circles, green squares and blue triangles;  photometric points shown as black circles 
were collected one rotation cycle before.  Fits with sine/cosine waves (plus first 
overtones for the left panels) are included (and shown as dashed lines) to outline the 
amount of variability caused by rotational modulation.  
$\pm1$~$\sigma$ error bars on data points are also shown whenever larger than symbols.  } 
\label{fig:var}
\end{figure*}

Emission in the \hei\ $D_3$ line at 587.562~nm, presumably produced in the post-shock region 
at the footpoints of accretion funnels, is considered as the most reliable accretion proxy.  
On V2129~Oph, \hei\ emission shows up as a narrow profile and amounts to an average 
equivalent width of about 11~\kms\ (i.e., 0.022~nm), with values ranging from 8 to 
19~\kms\ (see Fig.~\ref{fig:var} and top left panel of Fig.~\ref{fig:var}).  
Maximum \hei\ emission is found to occur at phase 
0.55, i.e., roughly coincident with times of minimum brightness (i.e., maximum spottedness);  
it suggests in particular that the accretion region traced by \hei\ emission is close to 
the cool spot at photospheric level.  Centred at an average RV of 
$0$~\kms, \hei\ emission is significantly red-shifted (by $\simeq7$~\kms) with respect to 
LSD photospheric profiles, indicating that it forms in a non-static region above the 
surface (e.g., the chromosphere) where the post-shock plasma is decelerating towards the star;  
the line also exhibits sinusoidal RV variations of about $\pm4$~\kms\ about the mean, 
suggesting that the corresponding emission concentrates in a region centred at phase 0.55 
(consistent with the observed maximum \hei\ emission) and presumably at high latitudes 
(to account for the small amplitude of the observed RV variations).  
At maximum emission, the emission flux is found to vary by almost a factor of 2 (from 
11 to 19~\kms, i.e., 0.022 to 0.037~nm) between successive rotation cycles (see Fig.~\ref{fig:var});  the 
dispersion is smaller at emission minimum (around 9~\kms, i.e., 0.018~nm).  During our previous run, 
\hei\ emission fluxes were grossly similar, varying from 6 to 19~\kms\ (i.e., 0.012 to 0.037~nm) 
across the rotation cycle.  

Clear Zeeman signatures are detected in conjunction with \hei\ $D_3$ emission (see Fig.~\ref{fig:bal} 
top right panel), tracing longitudinal fields ranging from 1.5 to 4.5~kG with magnetic maximum  
occurring at phase 0.55, i.e., at times of minimum brightness and maximum \hei\ 
emission (see Fig.~\ref{fig:var} second panel left column).  As usual for the \hei\ line 
\citep[e.g.,][]{Donati07, Donati08}, these Zeeman signatures feature a markedly non-antisymmetric 
shape (with a blue lobe stronger and narrower than the red one) indicating that the line is formed 
within a region featuring a strong velocity gradient, nicely supporting the idea that it traces 
non-static, chromospheric post-shock plasma decelerating towards the surface of the star.  
Note that longitudinal 
fields at similar phases but different rotation cycles grossly agree with each other, even 
when \hei\ emission fluxes (and thus presumably accretion rates) are significantly different 
(e.g., at cycles 1.609 and 2.542), indirectly confirming that \hei\ emission is probing  
localised regions with well defined field strengths and orientations at the surface of the star.  
These longitudinal field estimates are significantly 
larger than those estimated in the previous study \citep{Donati07}, 
by a typical factor of $2-3$;  it reveals in particular that the large-scale field of V2129~Oph
strongly varied on a timescale of only 4~yr.  
Maximum longitudinal fields from LSD and \hei\ profiles are found to occur at roughly the same 
phase in the present run, while they were shifted by about 0.25 cycle in our previous data set;  
this is additional confirmation that the large-scale field topology significantly changed between 
the two epochs.  

\begin{figure*}
\center{\hbox{\includegraphics[scale=0.6,angle=-90]{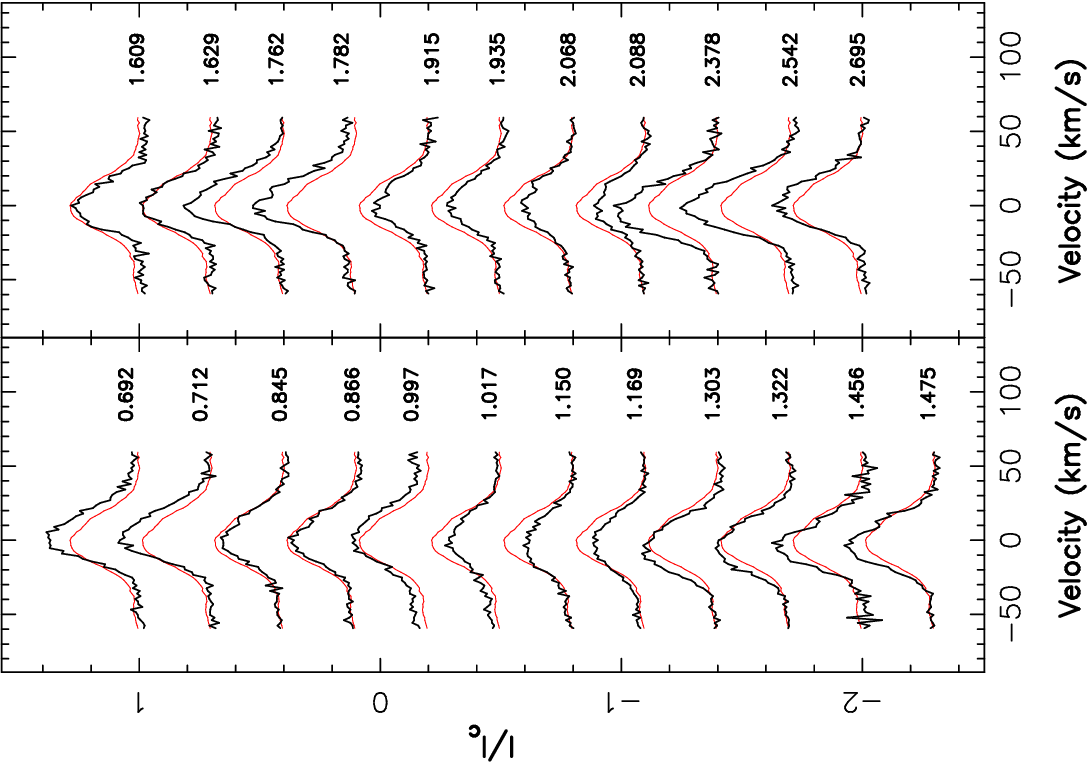}\hspace{4mm}
\includegraphics[scale=0.6,angle=-90]{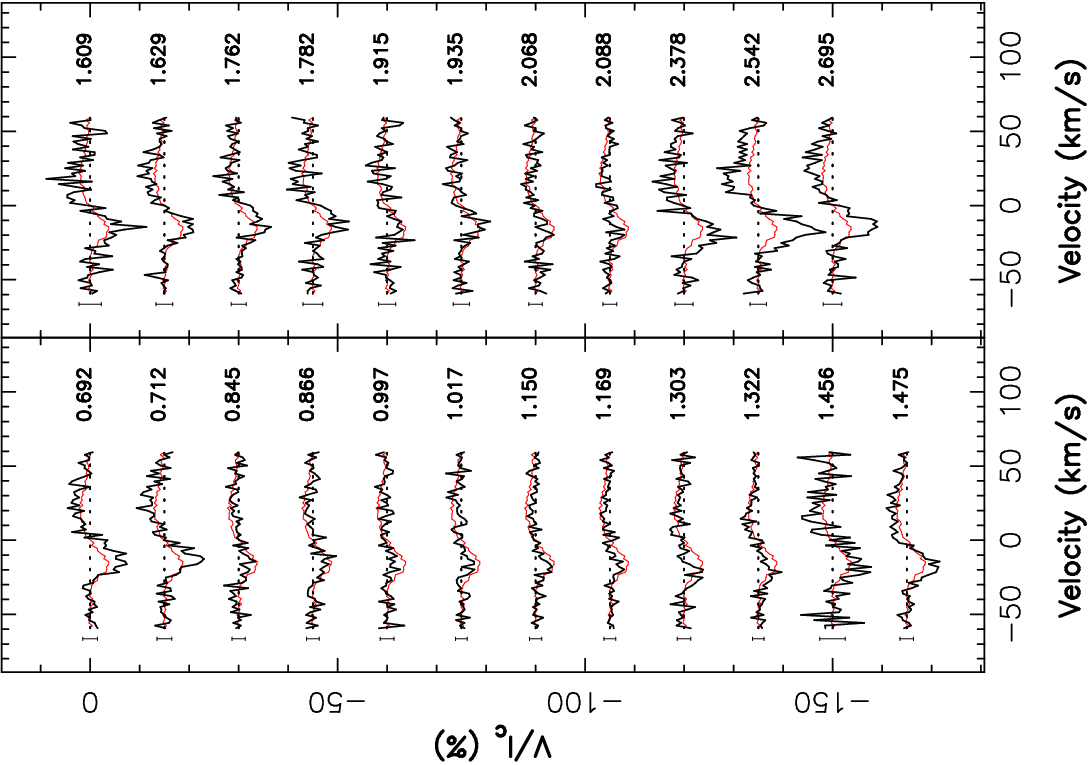}}}
\vspace{5mm}
\center{\hbox{\includegraphics[scale=0.6,angle=-90]{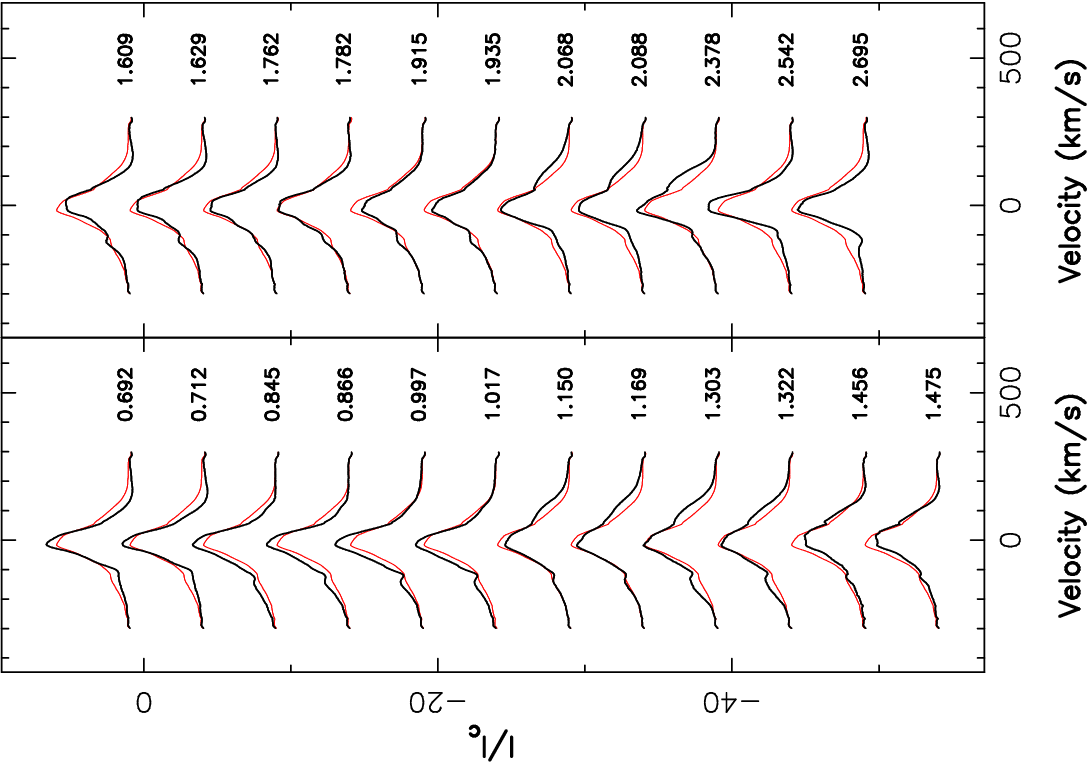}\hspace{4mm}
\includegraphics[scale=0.6,angle=-90]{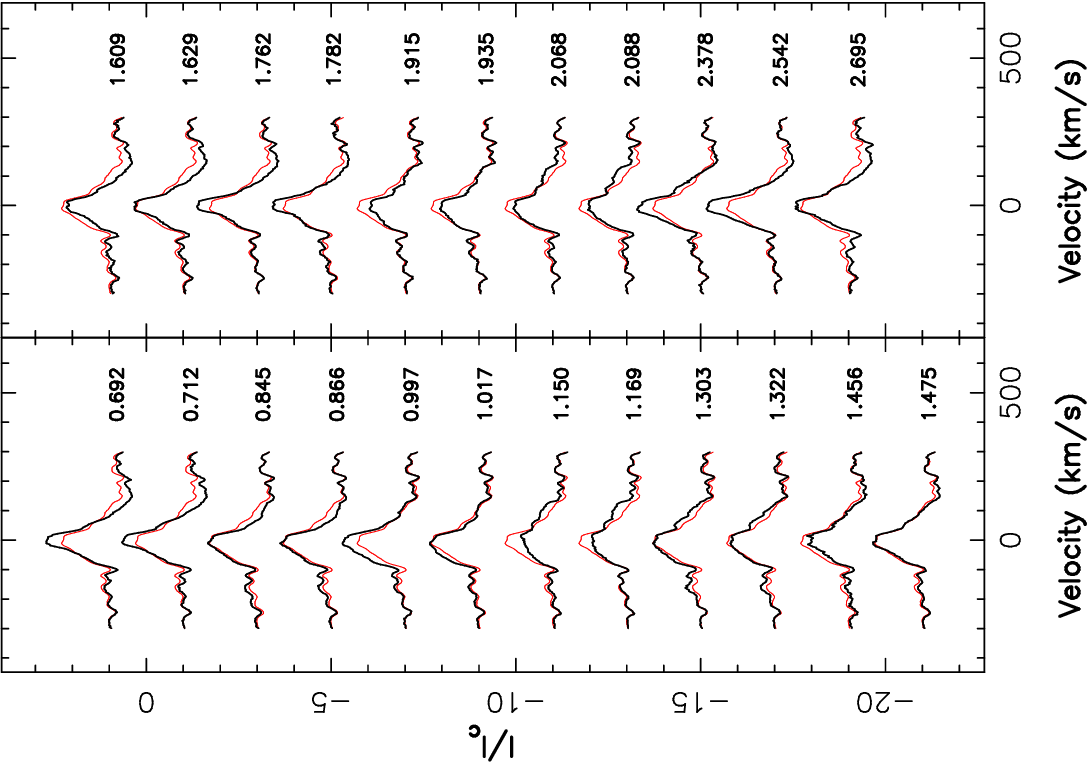}}}
\caption[]{Variation of the \hei\ $D_3$ (Stokes $I$: top left, Stokes $V$: top right),
H$\alpha$ (bottom left) and H$\beta$ (bottom right) lines of V2129~Oph in 2009 July.   
To emphasize variability, the average profile over the run is shown in red.  
Rotation cycles (as listed in Table~1) and 3$\sigma$ error bars (for Stokes $V$ data 
only) are also mentioned next to each profile.  }
\label{fig:bal}
\end{figure*}

\begin{figure*}
\center{\hbox{
\includegraphics[scale=0.37,angle=-90]{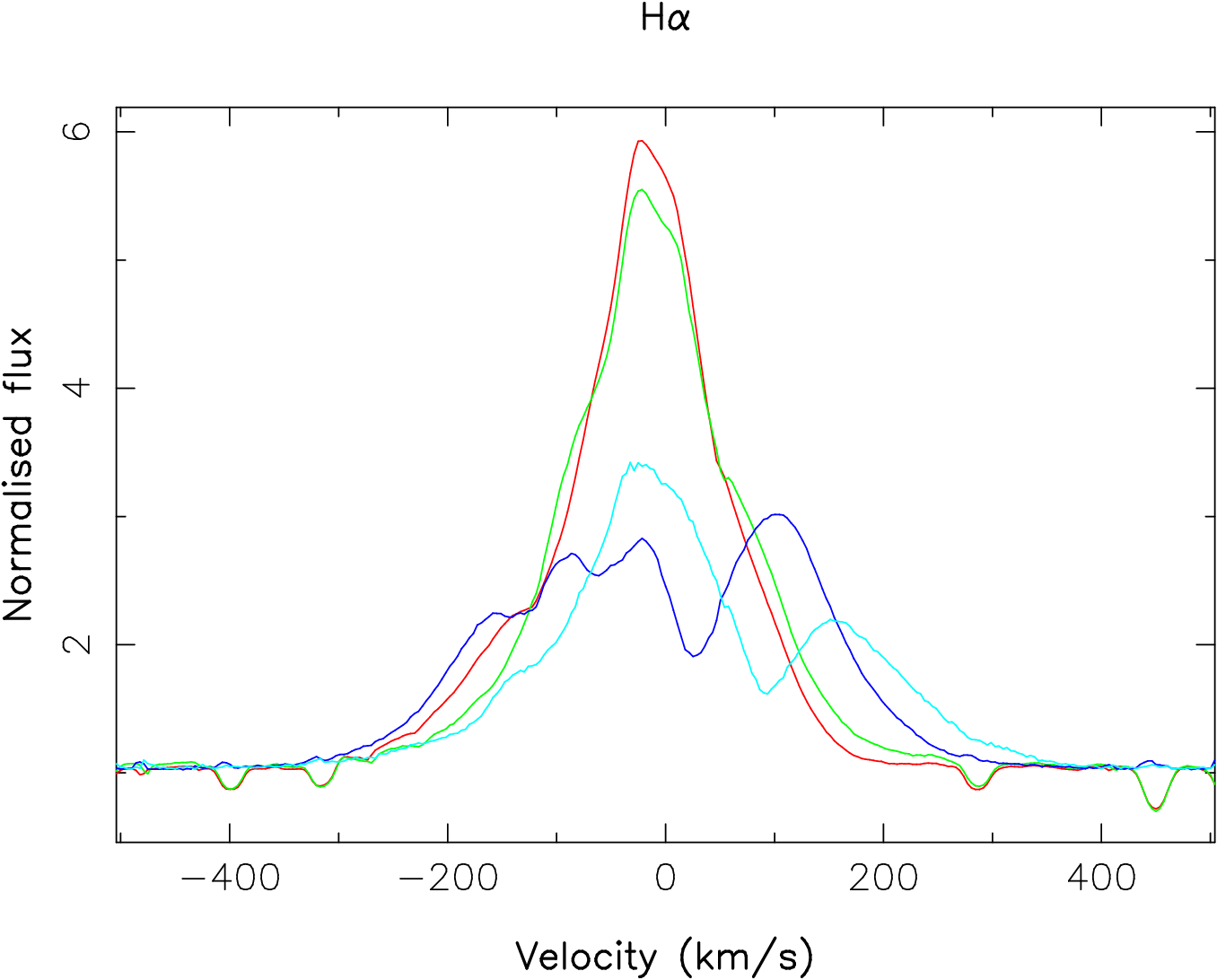}\hspace{5mm}
\includegraphics[scale=0.37,angle=-90]{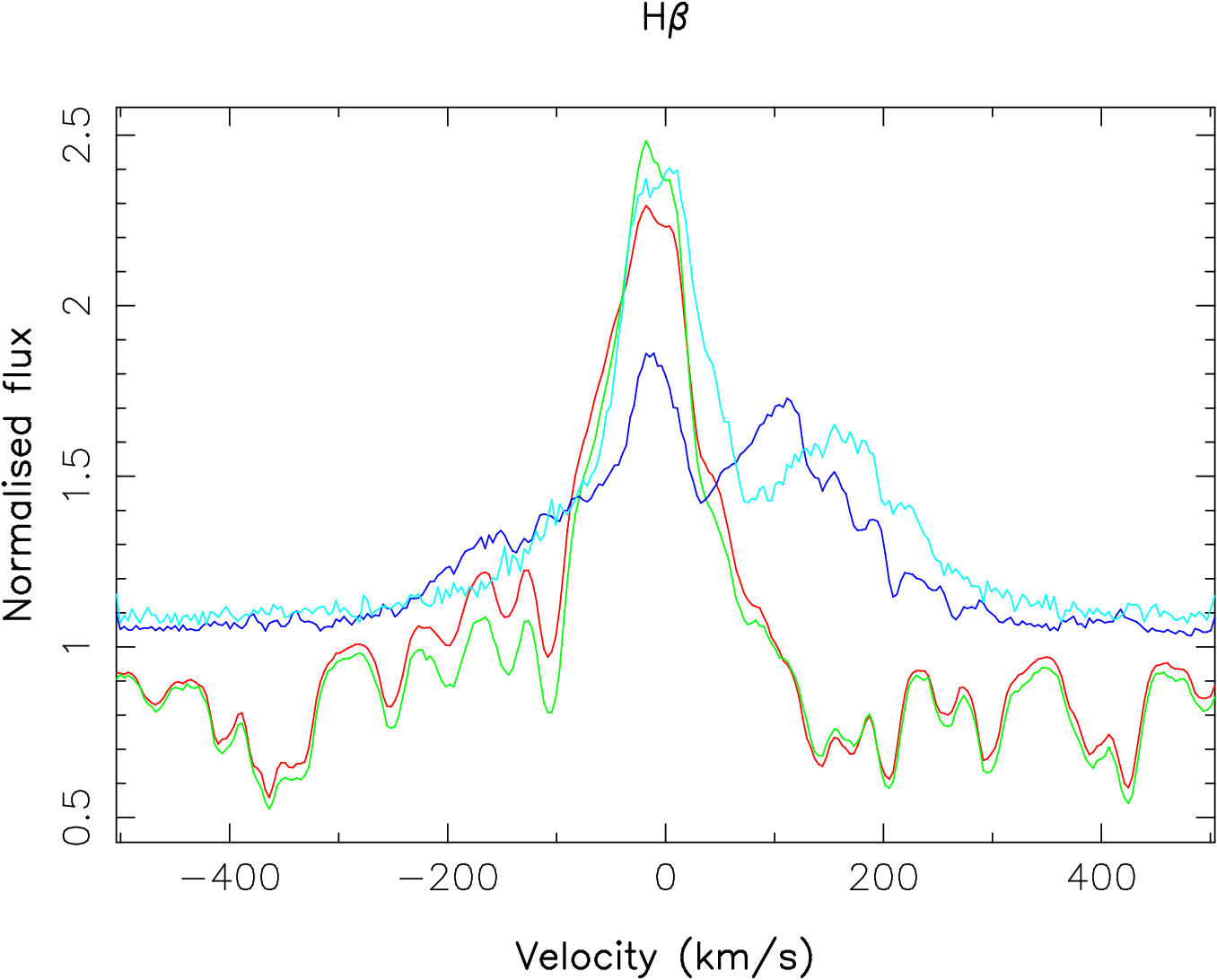}}}
\caption[]{Average (2009 July: red, 2005 June: green) 
and standard deviation (2009 July: dark blue, 2005 June: light blue) 
profiles of H$\alpha$ (left panel) and H$\beta$ (right panel) of V2129~Oph at both observing 
epochs.  Standard deviation profiles are expanded by a factor of 3 and shifted upwards by 1 for 
graphics purposes. }
\label{fig:bms}
\end{figure*}

Core emission in \caii\ IRT lines is another very useful probe of magnetospheric accretion, 
even though it often features a significant contribution from non-accreting chromospheres and 
is thus a more ambiguous proxy than \hei\ emission.  However, the redder spectral location, 
the higher magnetic sensitivity and the multiple nature of the corresponding spectral lines more than 
compensate for this drawback;  with their classical, nearly-antisymmetric profiles (with respect to 
the line center, see Fig.~\ref{fig:lsd} bottom right panel), the Zeeman signatures of \caii\ IRT lines 
are also simpler and easier to model than those of the \hei\ line.  
To extract the core \caii\ emission profiles (used later in the imaging analysis, see
Sec.~\ref{sec:mod}) from our spectra, we start by constructing a LSD-like weighted
average of the 3 IRT lines;  we then subtract the underlying (much wider) photospheric
absorption profile, with a Lorentzian fit to the far wings over a velocity interval of
$\pm200$~\kms\ about the emission core.
IRT emission is centred at a RV of about $-6$~\kms, i.e., redshifted by only 1~\kms\ with respect to 
LSD photospheric lines (as opposed to 7~\kms\ for the \hei\ line), confirming that it traces 
slowly-moving regions of the post-shock accretion funnels that are closer to the surface of the star.  
In 2009 July, equivalent widths of the \caii\ emission are in average 
equal to about 18~\kms\ (i.e., 0.050~nm) and vary from 12 to 24~\kms\ (i.e., 
$0.035-0.070$~nm) as a result of both rotational modulation and intrinsic variability (see 
Fig.~\ref{fig:var} right column third panel);  maximum IRT emission is found to occur at phase 
0.70, i.e., 0.15 rotation cycle later than \hei\ emission and magnetic maxima.  
RVs of IRT emission are found to vary by about $\pm2$~\kms\ about the mean, consistent with 
a localised emission region centred at phase 0.70 and at high latitudes.  

Clear Zeeman signatures are detected at all epochs in conjunction with \caii\ IRT emission, 
featuring peak-to-peak amplitudes of up to about 10\% (see Fig.~\ref{fig:lsd} bottom right panel).  
Corresponding longitudinal fields vary from 0.4 to 1.2~kG (see Fig.~\ref{fig:var} left column 
third panel);  this is about twice larger than in our previous data set \citep{Donati07}, 
confirming again that the field has considerably strengthened within the corresponding 4~yr.  
Maximum longitudinal field is observed at phase 0.55, i.e., in conjunction with the 
magnetic maximum traced by \hei\ lines.  The longitudinal field curve is found to repeat 
fairly well over the two rotation cycles, clearly demonstrating that rotational modulation 
is dominant.  The ratio of \hei\ to \caii\ longitudinal fields is $\simeq4$ across the whole 
run;  this essentially reflects that \caii\ longitudinal fields suffer from dilution and flux 
cancellation (with respect to those from \hei) as a result of \caii\ emission forming over a 
much wider area (including in particular the quiet chromosphere) while \hei\ emission traces 
only the accretion regions.  

As usual for cTTSs, Balmer lines (and in particular H$\alpha$ and H$\beta$) are in emission and 
show clear temporal variability and rotational modulation (see Fig.~\ref{fig:bal} lower panel).  
As in our previous data set, H$\alpha$ and H$\beta$ mostly show one main emission peak centred 
at about -15~\kms, blue-shifted by about 8~\kms\ with respect to the LSD photospheric profiles;  
the average equivalent widths of both profiles are equal to 800 and 150~\kms\ respectively 
(i.e., 1.7 and 0.24~nm), i.e., very similar to those measured in 2005 June.  
Looking at the variance profiles of both lines (see Fig.~\ref{fig:bms}), one can see that temporal 
variations mostly affect the central emission peak and the red wing, both parts of the line 
being anti-correlated with each other;  this is similar to what was reported from our previous 
data set, except that variations in the central peak are now smaller than in 2005 June while those in 
the red wing are both larger and shifted to smaller velocities (with the variance profile now 
peaking at 100~\kms\ instead of 150~\kms).  In our new data set, Balmer emission and profile 
variations are also clearly visible in the blue wing (around $-170$~\kms), both in H$\alpha$ 
and H$\beta$ (see Fig.~\ref{fig:bms});  this emission component was not detected in the 
previous data set.  

Recurrent absorption episodes in the red wing of Balmer lines are usually interpreted 
as evidence for accretion veils anchored in the inner regions of the surrounding disc and 
periodically intersecting the line of sight as the star rotates \citep[e.g.,][]{Bouvier07}.  
Those observed on V2129~Oph in 2009 July occur between phases 0.6 and 0.9 (see 
Fig.~\ref{fig:bal} lower panel), i.e., $0.05-0.35$ rotation cycle (or 0.20 cycle in 
average) after \hei\ and magnetic maxima;  it suggests that the corresponding accretion veils 
are trailing the main magnetic spot detected at the surface of the protostar (centred at phase 
0.55).  In 2005 June, these absorption episodes were shorter, being visible for less than 15\% 
of a rotation cycle and occurring slightly (about 0.05 rotation 
cycle) before magnetic maximum \citep{Donati07}.  

From the average equivalents widths of the \hei, \caii\ IRT, H$\beta$ and H$\alpha$ lines, we 
derive logarithmic line fluxes (with respect to the luminosity of the Sun \lsun) equal to $-5.1$, 
$-4.8$, $-4.1$ and $-3.2$ respectively\footnote{To derive line fluxes from normalized equivalent 
widths, we approximate the continuum level by a Planck function at the temperature of the stellar 
photosphere.  Results are found to be compatible with those in the published literature 
\citep[e.g.,][]{Mohanty05} within better than 0.1~dex.  }, implying logarithmic accretion 
luminosities (with respect to \lsun) of $-2.0$, $-2.0$, $-2.3$ and $-1.7$ respectively 
\citep[using empirical correlations from][]{Fang09}.  As estimates from the two main accretion proxies 
agree with each other and with the overall mean, we can safely conclude that the average 
logarithmic accretion luminosity of V2129~Oph is $-2.0\pm0.3$ and thus that the average 
logarithmic mass accretion rate (in \mspy) is equal to $-9.2\pm0.3$.  Using the same relations, 
we find that the logarithmic accretion rate at magnetic maximum is typically $-9.0$ (with 
peaks of up to $-8.9$) and about $-9.4$ around magnetic minimum (i.e., phase 0.05), the observed 
fluctuation most likely tracing the varying viewing configuration (rather than an intrinsic 
change in the accretion rate).  
Mass accretion rates can independently be estimated through the full width of H$\alpha$ at 
10\% height \citep[e.g.,][]{Natta04, Cieza10};  in the case of V2129~Oph in 2009 July, 
H$\alpha$ exhibits a full width of 350~\kms\ implying an logarithmic mass accretion rate estimate 
of $-9.4\pm0.6$ in reasonably good agreement with our main estimate.  

Line equivalent widths and fluxes measured during our 2005 run (and in particular those of 
\hei\ and \caii\ IRT accretion proxies) are basically identical to those listed here, suggesting 
that the mass accretion rate at the surface of V2129~Oph is roughly stable (within a factor of 
$\simeq$2) on long timescales as well;  in particular, this is different 
than the behaviour reported for AA~Tau, where the mass accretion rate can vary by an order 
of magnitude between successive rotation cycles \citep{Donati10b}.  Further data collected 
over a longer time span are however needed to confirm that we did not catch V2129~Oph in an 
unusually quiet state of accretion at both epochs.

\section{Modelling the surface of V2129~Oph}
\label{sec:mod}

\subsection{Overview of the modelling method}

Using the set of LSD and \caii\ IRT profiles of V2129~Oph described in the previous section, we can 
recover the large-scale field topology at the surface of the protostar, as well as maps of how 
photospheric brightness and accretion-powered \caii\ emission distributes with longitude and 
latitude.  In this aim, we apply our new modelling technique, detailed extensively and proved 
successful in a previous MaPP study \citep{Donati10b}.  In the present paper, 
we only briefly describe the technical aspects of this method and concentrate on its specific 
application to our new V2129~Oph data set;  we also carry out a new analysis of the older data 
set \citep[collected in 2005 June and analysed with an experimental version of our imaging 
tool,][]{Donati07} to allow a more direct and meaningful comparison of reconstructed maps.  

Following the principles of maximum entropy, our code automatically derives the simplest magnetic
topology, photospheric brightness image and accretion-powered \caii\ emission map compatible with 
the series of rotationally modulated Stokes $I$ and $V$ LSD and \caii\ IRT profiles.
The reconstruction process is iterative and proceeds by comparing at each step the synthetic
Stokes $I$ and $V$ profiles corresponding to the current images with those of the observed data set.
The magnetic field is described through its poloidal and toroidal components expressed
as spherical-harmonics (SH) expansions \citep{Donati06b}.  The spatial distribution of photospheric
brightness (with respect to the quiet photosphere) and that of accretion-powered \caii\ emission (in 
excess of and with respect to that produced by the quiet chromosphere) are modelled as series of 
independent pixels (typically a few thousand) on a grid covering the visible surface of the star 
(with spots in the brightness image assumed to be darker/cooler than the quiet photosphere and spots 
in the accretion-powered \caii\ emission map supposed to be brighter than the quiet chromosphere).

Synthetic profiles are computed by summing up the elementary spectral contributions from each image
pixel over the visible hemisphere, taking into account all relevant local parameters of the
corresponding grid cell (e.g., brightness, accretion-powered excess emission, magnetic field strength
and orientation, radial velocity, limb angle, projected area).  Since the problem is partly ill-posed,
we stabilise the inversion process by using an entropy criterion (applied to the SH
coefficients and to the brightness/excess emission image pixels) aimed at selecting the field
topologies and images with minimum information among all those compatible with the data.
The relative weights attributed to the various SH modes can be imposed, e.g., for purposedly
producing antisymmetric or symmetric field topologies with respect to the centre of the star
\citep[by favouring odd or even SH modes,][]{Donati07, Donati08}.
More details concerning the specific description of local profiles used in the model can be 
found in \citet{Donati10b}.  

\begin{figure*}
\center{\hbox{
\includegraphics[scale=0.37,angle=-90]{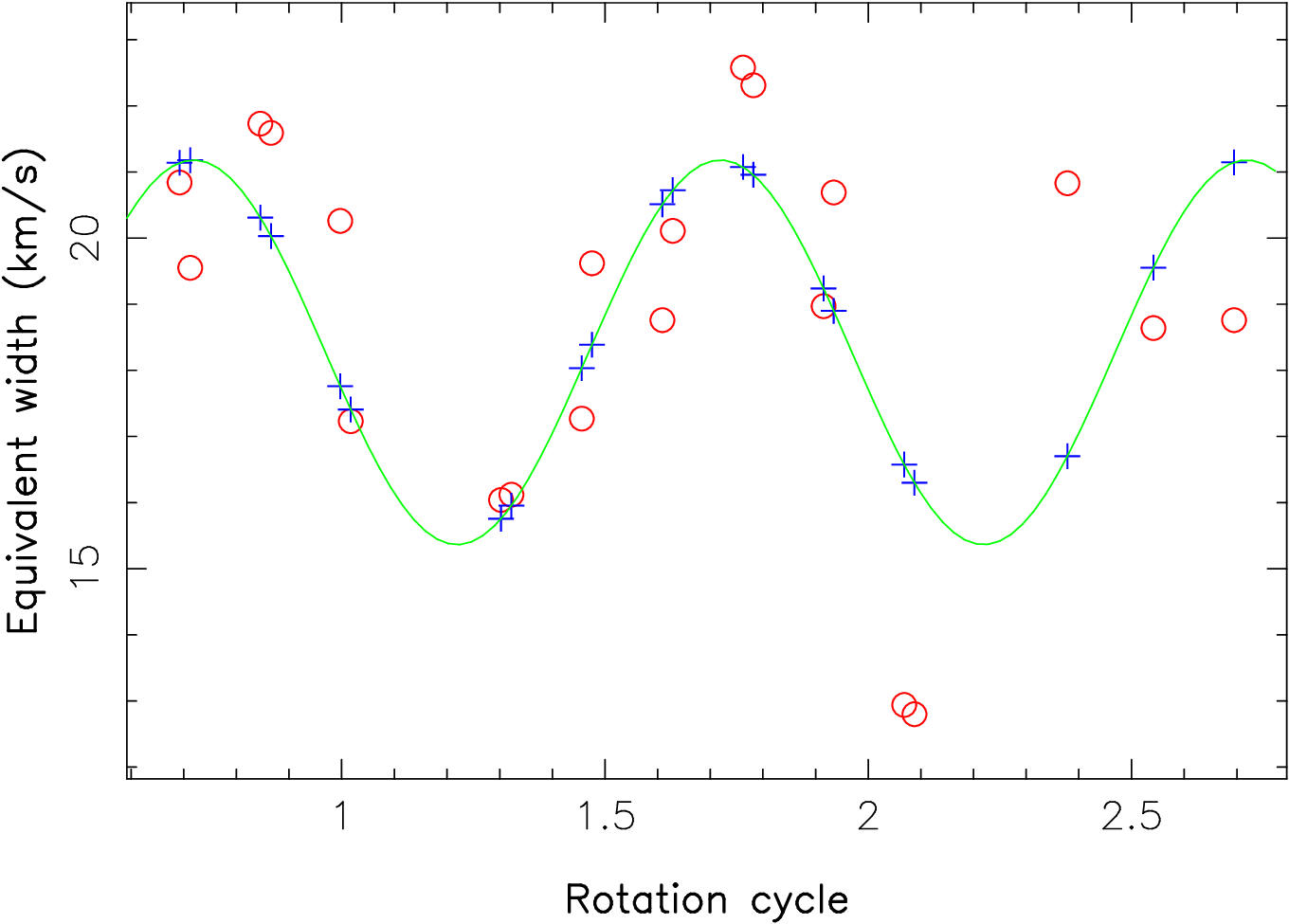}\hspace{5mm}
\includegraphics[scale=0.37,angle=-90]{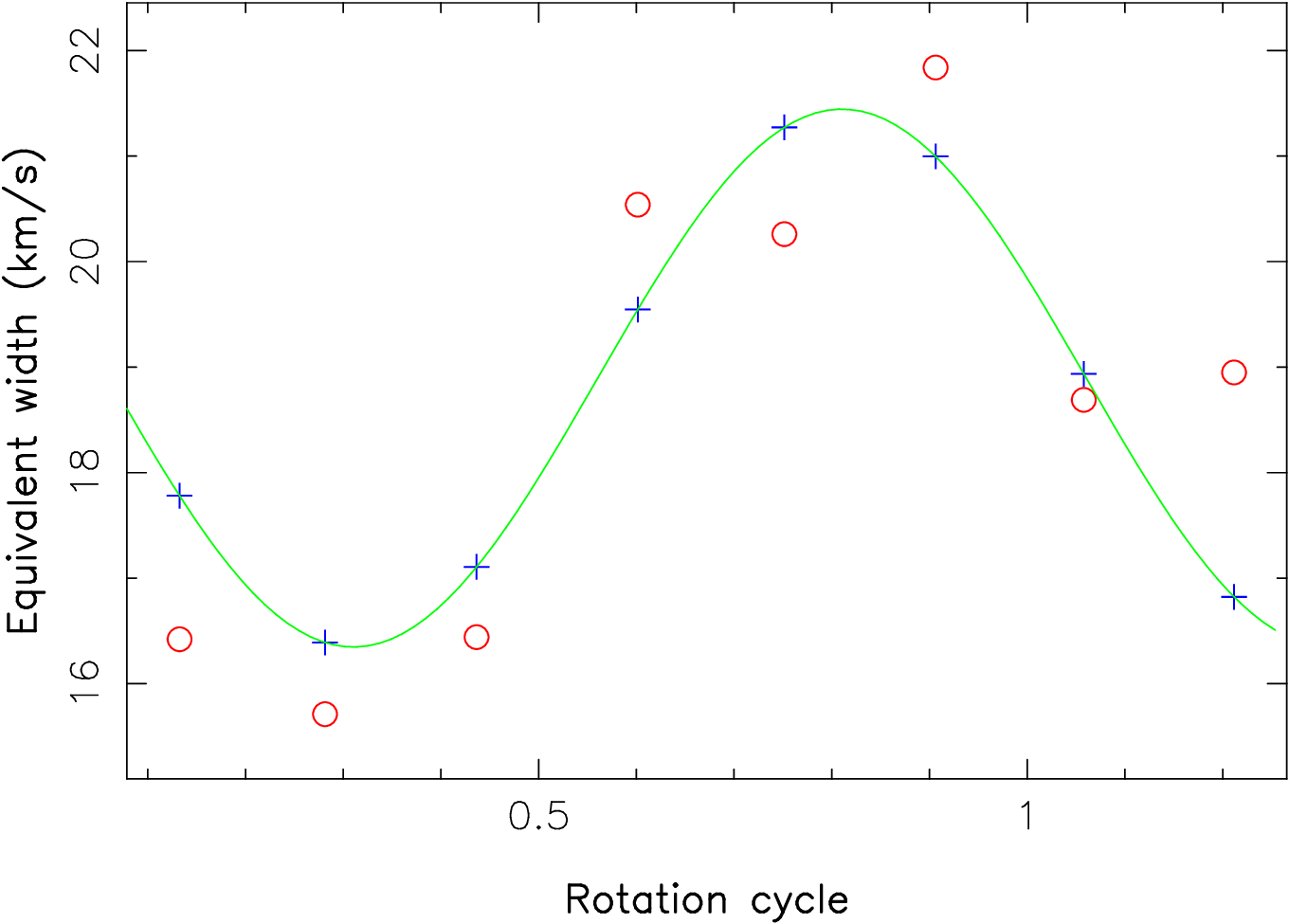}}}
\caption[]{Measured (red open circles) and fitted (blue pluses) equivalent widths of
the \caii\ IRT LSD profiles of V2129~Oph in 2009 July (left panel) and 2005 June (right panel).  
The model wave (green line) providing the best (sine+cosine) fit to 
the data (with a period of 6.53~d) presumably traces rotational modulation, while the 
deviation from the fit illustrates the strength of intrinsic variability.  }
\label{fig:ew}
\end{figure*}

\subsection{Application to V2129~Oph}

Our imaging model assumes that the observed profile variations are mainly due to rotational modulation, 
and potentially to surface differential rotation as well when the star is observed for at least several 
rotation cycles;  all other sources of profile variability (and in particular intrinsic variability) 
cannot be properly reproduced and thus contribute as noise into the modelling process, degrading the
imaging performance and potentially even drowning all relevant information.
Filtering out significant intrinsic variability from the observed profiles is thus worthwhile to 
optimise the behaviour and convergence of the imaging code.  

We implement this by applying specific corrections on our data set.  We first suppress veiling
by scaling all LSD Stokes $I$ and $V$ photospheric profiles, to ensure that unpolarized lines have 
the same equivalent widths.  We also remove the non rotationally-modulated part in the observed 
fluctuations of \caii\ IRT emission, by fitting them with a sine+cosine wave (see Fig.~\ref{fig:ew}) 
and by scaling the
corresponding Stokes $I$ and $V$ profiles accordingly, thus ensuring that equivalent widths 
of unpolarized lines match the optimal fit.  Although only approximate (especially the removal of 
the intrinsic variability), this procedure is at least very straightforward and has proved successful 
when applied to real data \citep[e.g.,][]{Donati10b} and efficient at retaining rotational modulation 
mostly.  

\begin{figure*}
\vspace{-2mm}
\hbox{\includegraphics[scale=0.7]{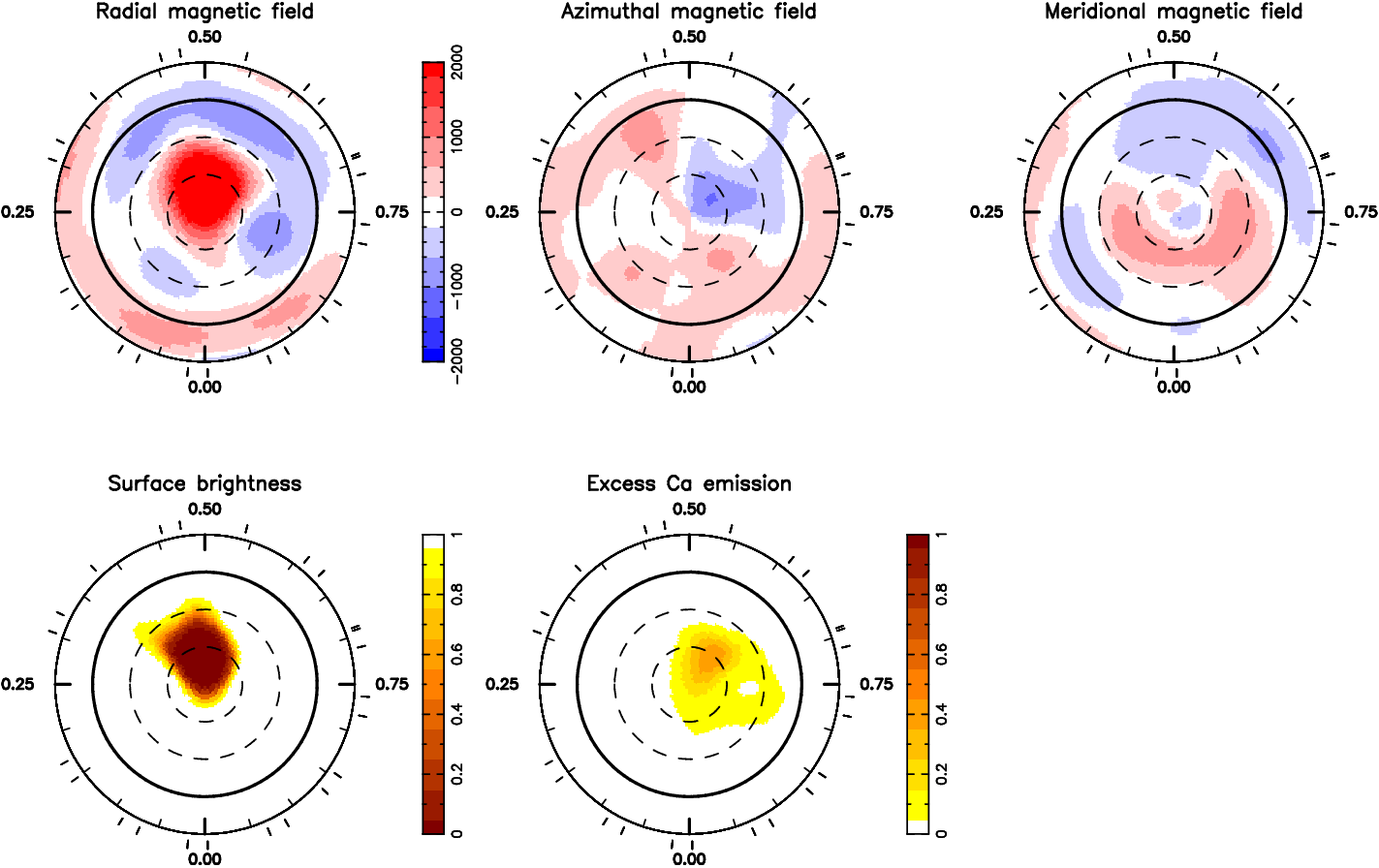}}
\vspace{8mm}
\hbox{\includegraphics[scale=0.7]{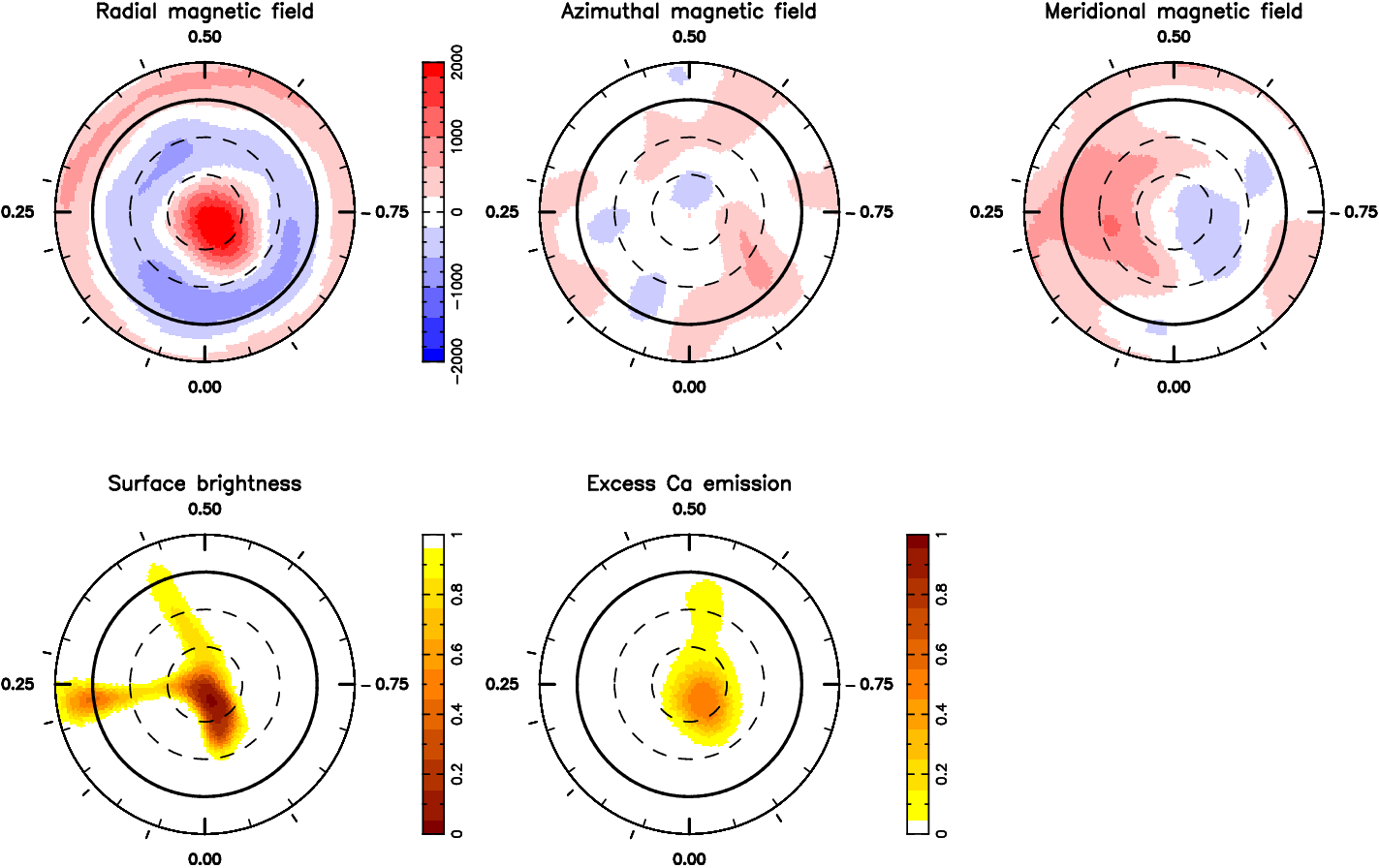}} 
\caption[]{Maps of the radial, azimuthal and meridional components of the magnetic field $\bf B$ 
(first and third rows, left to right panels respectively), photospheric brightness and excess 
\caii\ IRT emission (second and fourth rows, first and second panels respectively) at the 
surface of V2129~Oph in 2009 July (two upper rows) and 2005 June (two lower rows).  
Magnetic fluxes are labelled in G;  local photospheric brightness (normalized to that of the quiet 
photosphere) varies from 1 (no spot) to 0 (no light);  local excess \caii\ emission varies from 0 
(no excess emission) to 1 (excess emission covering 100\% of the local grid cell, assuming an  
intrinsic excess emission of 10$\times$ the quiet chromospheric emission).  
In all panels, the star is shown in flattened polar projection down to latitudes of $-30\degr$, 
with the equator depicted as a bold circle and parallels as dashed circles.  Radial ticks around 
each plot indicate phases of observations. } 
\label{fig:map}
\end{figure*}

\begin{figure*}
\center{\hbox{
\includegraphics[scale=0.6,angle=-90]{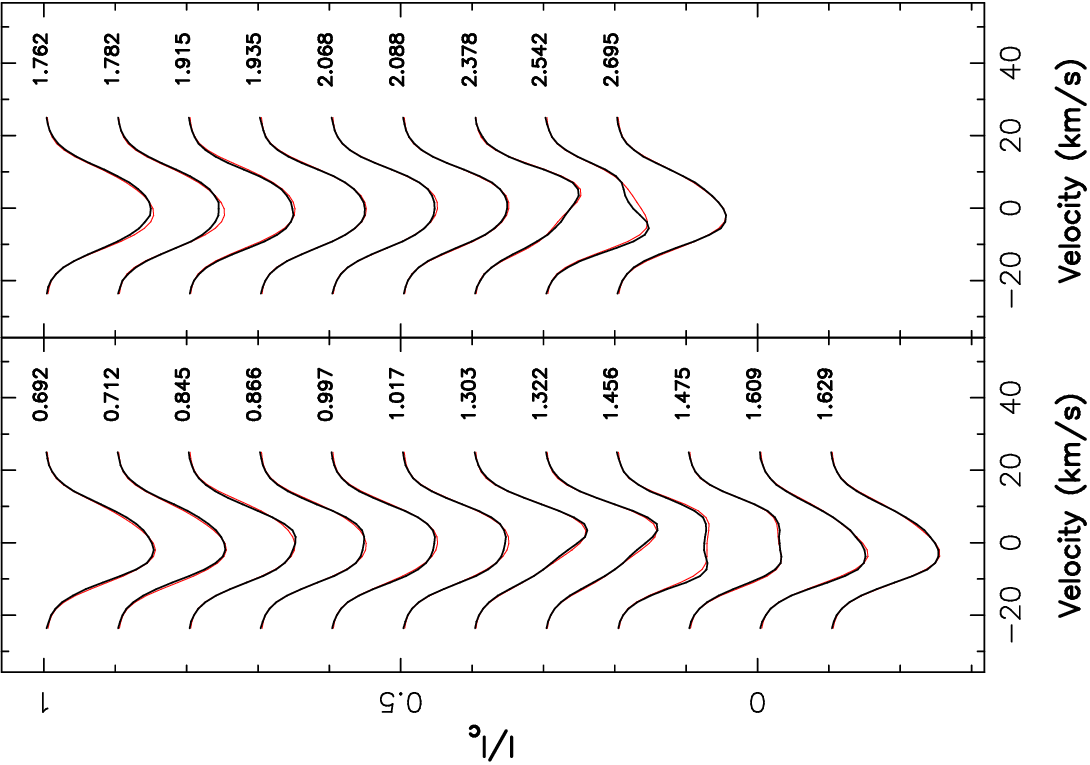}\hspace{4mm}
\includegraphics[scale=0.6,angle=-90]{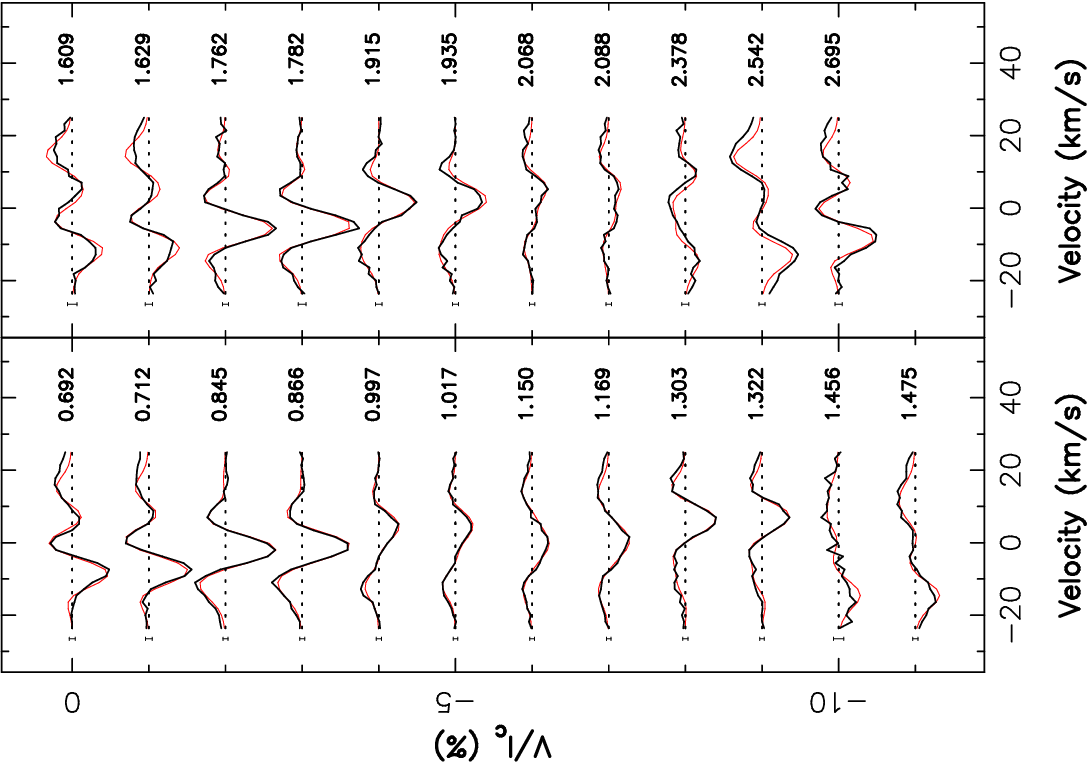}}}  
\vspace{6mm}
\center{\hbox{
\includegraphics[scale=0.6,angle=-90]{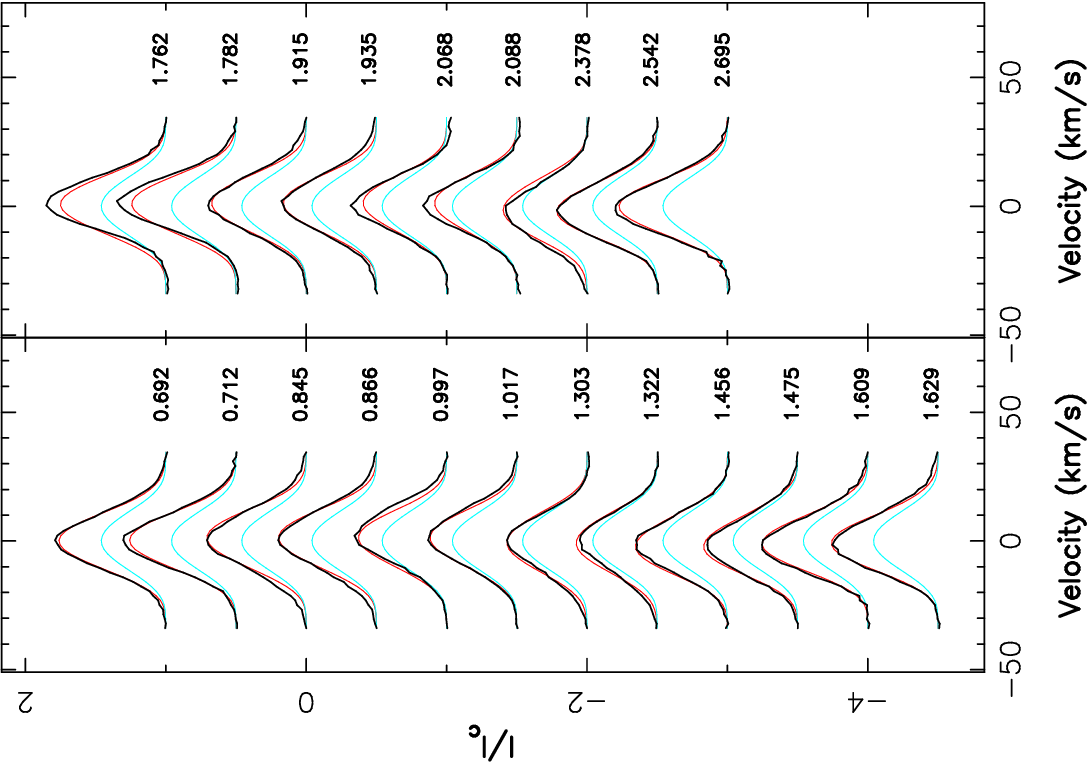}\hspace{4mm}
\includegraphics[scale=0.6,angle=-90]{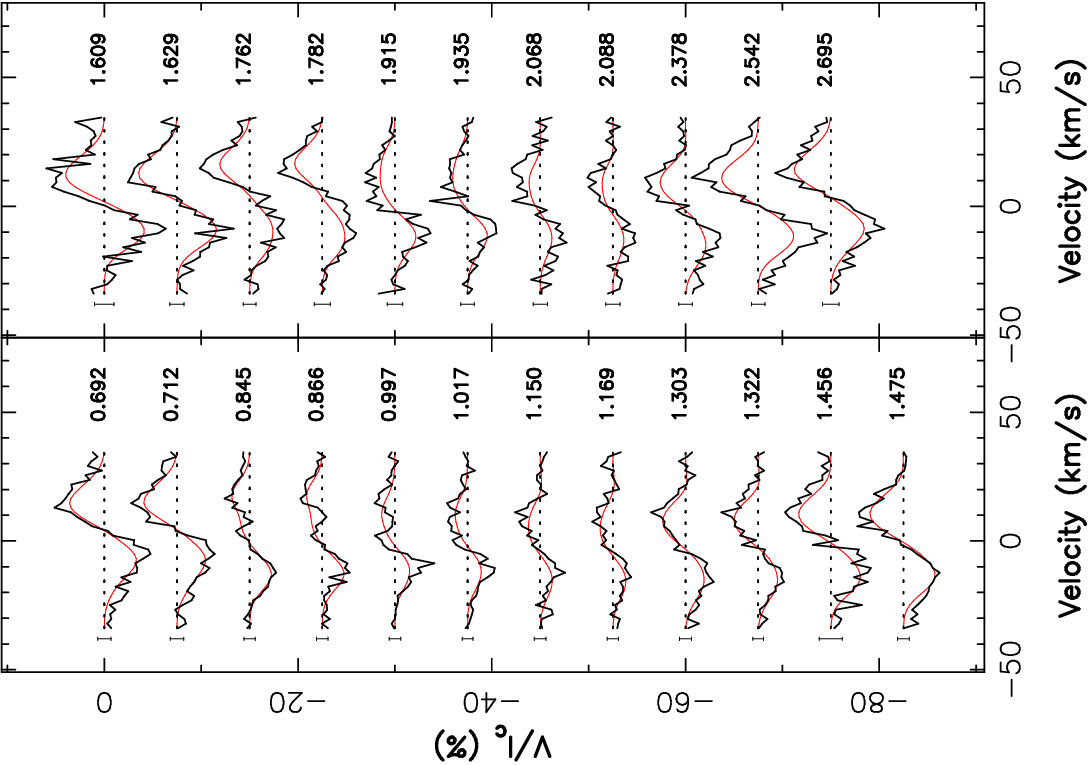}}} 
\caption[]{Maximum-entropy fit (thin red line) to the observed (thick black line) Stokes $I$ and 
Stokes $V$ LSD photospheric profiles (first two panels) and \caii\ IRT profiles (last two panels) 
of V2129~Oph in 2009 July.  The light-blue curve in the bottom left panel shows the (constant) 
contribution of the quiet chromosphere to the Stokes $I$ \caii\ profiles.
Rotation cycles and 3$\sigma$ error bars (for Stokes $V$ profiles) are also shown next to each 
profile. } 
\label{fig:fit}
\end{figure*}

\begin{figure*}
\center{\hbox{\includegraphics[scale=0.57,angle=-90]{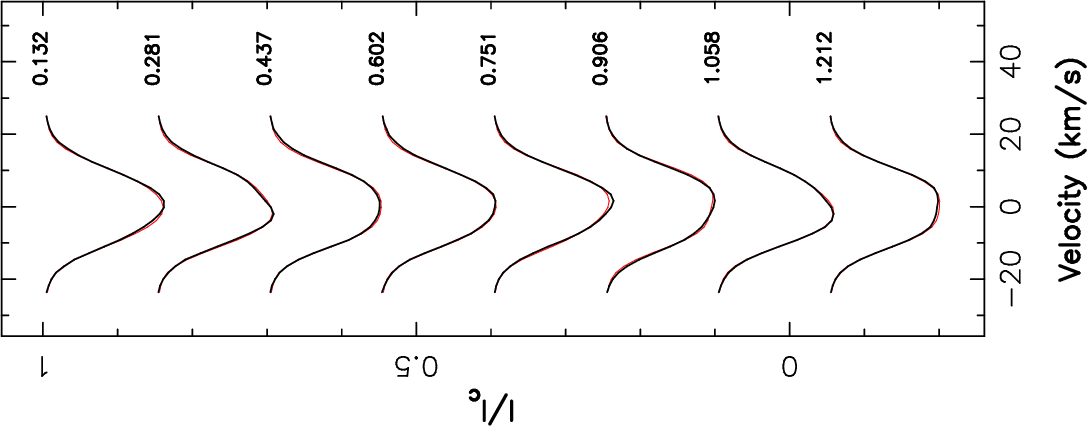}\hspace{3mm}
\includegraphics[scale=0.57,angle=-90]{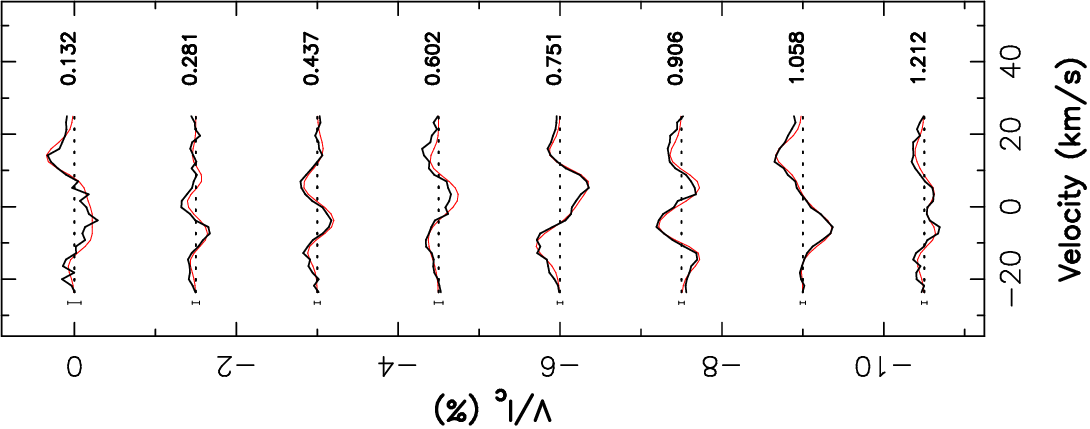}\hspace{3mm}
\includegraphics[scale=0.57,angle=-90]{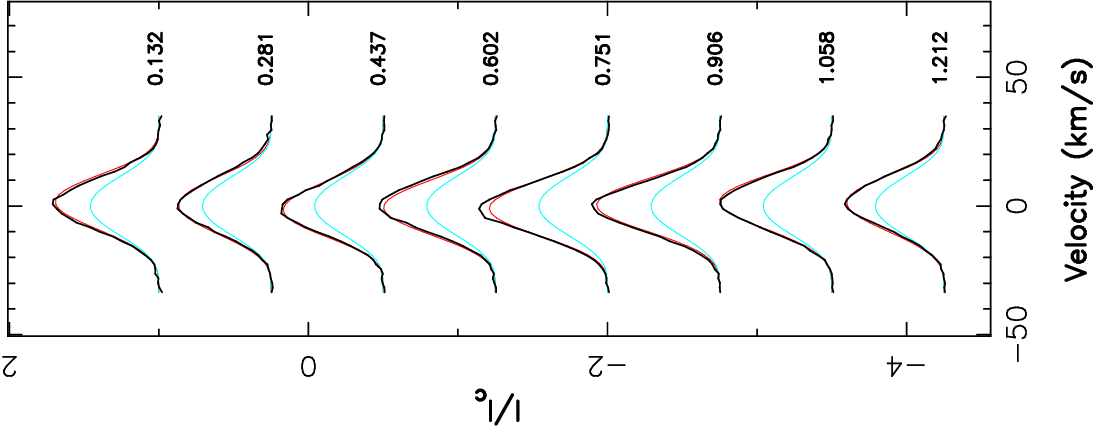}\hspace{3mm}
\includegraphics[scale=0.57,angle=-90]{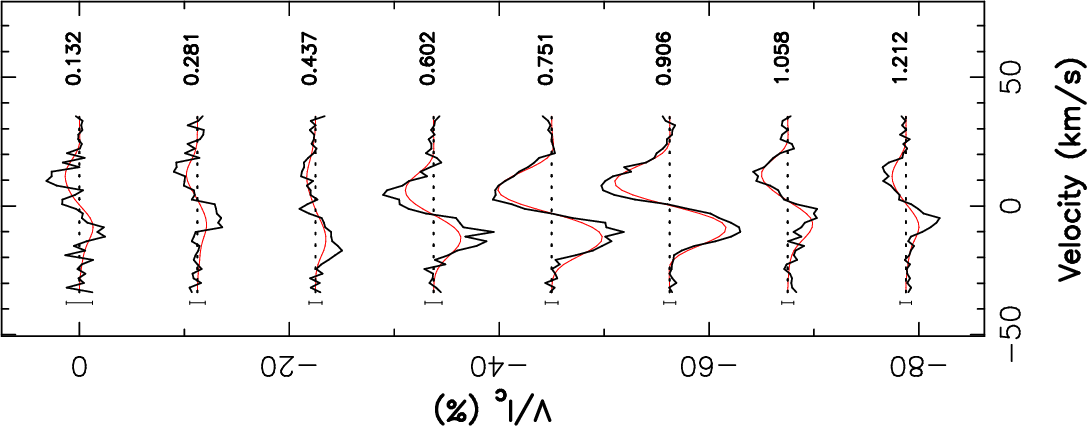}}}
\caption[]{Same as Fig.~\ref{fig:fit} for the 2005 June data set. } 
\label{fig:fit2}
\end{figure*}

The reconstructed magnetic, brightness and accretion maps of V2129~Oph are shown in 
Fig.~\ref{fig:map}, with corresponding fits to the data shown in Figs.~\ref{fig:fit} and 
\ref{fig:fit2};  we also include new maps (and fits) derived from the previous data set, 
but reconstructed with our new modelling method to ease comparison between the two sets 
of images.  
The SH expansion describing the field 
was limited to terms with $\ell\leq8$, very small power being reconstructed in higher 
order modes.  
Both images were reconstructed assuming that the field is antisymmetric with 
respect to the centre of the star (see below).  

The overall fits to the data are reasonably good, reproducing the data down to reduced 
chi-squares \chisqr\ of about 2 (starting from initial \chisqr\ of 25 and 40 for the 
2009 July and 2005 June data sets respectively).  
Optimal fits are obtained for $\vsini=14.5\pm0.5$~\kms\ and $\vrad=-7.0\pm0.1$~\kms; 
\caii\ IRT lines are found to be slightly redshifted (by about 1~\kms) with respect to 
photospheric lines, as usual for cTTSs.  
We also find that optimal fits to the observed 
Stokes $V$ profiles can be achieved when setting the local filling factor $\psi$, describing 
the relative proportion of magnetic areas at any given point of the stellar surface, to 
$\psi\simeq0.4\pm0.1$.  
The emission profile scaling factor $\epsilon$, describing the emission enhancement of 
accretion regions over the quiet chromosphere, is again set to $\epsilon=10$ \citep[as 
in][]{Donati10b}, yielding fractional areas of accretion spots of about $2-3$\% at 
both epochs.  Most other model parameters, and in particular those describing the local 
line profiles, are either identical of very similar to those chosen in our previous studies;  
for more information, the reader is referred to \citet{Donati10b}.  

\subsection{Results}

The reconstructed large-scale magnetic topology of V2129~Oph in 2009 July is 
mostly poloidal, with only 5\% of the magnetic energy in the toroidal component;  
this poloidal component is mainly antisymmetric, with at least $\simeq60$\% of the energy 
concentrating in modes with $m<\ell/2$.  We find that the field is more likely to be 
antisymmetric about the centre of the star, a symmetric topology requiring much more 
magnetic energy to produce the same fit to the data.  
The dominant terms in the reconstructed SH expansion of the large-scale poloidal field 
correspond to $\ell=3$ (with $m=0,1$, i.e., a moderately tilted octupole) and $\ell=1$ 
(with $m=0$, i.e., an aligned dipole), whose respective intensity at the pole are about 
$2.1\pm0.3$~kG and $0.9\pm0.1$~kG respectively;  
from different reconstructions with slightly different input parameters and reconstruction 
schemes, we estimate that the octupole is typically $2-3$ stronger than the dipole.  
The main magnetic feature that we reconstruct is a high-latitude radial field region where 
the magnetic intensity peaks at about 4~kG, slightly smaller though comparable to the maximum 
fields traced by the \hei\ line (about 5~kG, see Fig.~\ref{fig:var});  this main magnetic 
region is centred at phase 0.5, compatible with the estimate guessed in Sec.~\ref{sec:var} 
from the longitudinal field curves of both \caii\ IRT and \hei\ lines.  The octupole 
component of the poloidal field can be directly traced in the reconstructed magnetic map, 
as a low-latitude ring of negative radial field encircling the main high-latitude radial 
field region. 

The reconstructed magnetic maps of V2129~Oph in 2005 June are grossly similar to 
those of 2009 July except for the large change in the overall field intensity;  they are also 
compatible with the preliminary version of \citet{Donati07}.  Again, the large-scale field 
is found to be mostly poloidal and axisymmetric, with the poloidal component being 
dominated by a moderately tilted octupole (with a polar strength of about $1.5\pm0.3$~kG);  
the dipole component is typically 5 times smaller than the octupole component, with 
polar strengths of about 0.3~kG.  The high-latitude radial field region features peak 
intensities of about 2~kG, slightly larger though compatible with fields traced by 
the \hei\ line.  The phase at which this high field region is located is different in 
2009 July and 2005 June;  however, this difference likely reflects no more than the 
uncertainty on the rotation period.  
The reconstructed maps thus confirm the preliminary conclusions of Sec.~\ref{sec:var} 
about the overall strengthening of the field (by typically a factor of 2) between 2005 and 
2009;  they further suggest that this increase mostly concerns the dipole component of the 
field (about thrice stronger in 2009) and to a lesser extent its octupole component 
(about 50\% stronger in 2009).  

The photospheric brightness map we reconstruct for V2129~Oph in 2009 July essentially 
features one large dark spot at high latitude covering about 6.5\% of the total stellar 
surface (i.e., about 25\% of the visible hemisphere at maximum visibility) and within most 
of which the brightness contrast at visible wavelengths (with respect to the surrounding 
photosphere) is larger than 20;  this dark spot mostly overlaps the strongly magnetic 
region discussed above, making the latter hardly visible at optical wavelengths (hence 
its non-detection in LSD profiles and in the corresponding longitudinal field curve, see 
Fig.~\ref{fig:var}).  This photospheric brightness map is also grossly consistent (both 
in phase and amplitude) with the observed photometric light-curve of V2129~Oph, even 
though the latter was not used in the modelling.  
The photospheric brightness map corresponding to 2005 June \citep[again compatible with, 
though less detailed than, the previous version of][]{Donati07} features a smaller and 
less contrasted high-latitude dark spot as well as low-latitude appendages, covering 
altogether about 5\% of the total stellar surface.  As in 2009 July, the high-latitude 
spot is mostly coincident with the strongly magnetic region of the radial field map.  
We estimate that the increase in size and contrast of the high-latitude dark spot on 
V2129~Oph between 2005 and 2009 is real, and readily traceable in the comparatively 
larger distortions (by about a factor of 2 in average, as judged from standard deviation 
profiles) in the unpolarized LSD profiles of the 2009 data set (see first panels of 
Figs.~\ref{fig:fit} and \ref{fig:fit2}).  

The map of accretion-powered excess emission that we reconstruct from the 2009 July data 
set shows one main high-latitude region, covering about 2.5\% of the total stellar 
surface.  This region is located at phase 0.6, i.e., slightly trailing by about 0.1 cycle 
the main radial field region and the dark photospheric spot;  this phase shift
is readily traceable in the rotational modulation of both the equivalent widths and radial 
velocities of the \caii\ IRT emission (see Sec.~\ref{sec:var}), suggesting that it is real 
even though corresponding to a spatial size comparable to our resolution element at the 
surface of the star.  
A similar accretion region is reconstructed from the 2005 June data set, this time 
overlapping the main field region and dark photospheric spot.  

Our modelling can also estimate the amount by which the photosphere is sheared by 
differential rotation \citep[e.g.,][]{Donati03b, Donati10}.  This is done by assuming that 
the rotation rate at the surface of the star is varying with latitude $\theta$ (as 
$\omeq - \dom \sin^2 \theta$ where \omeq\ is the rotation rate at the equator \dom\ the 
difference in rotation rate between the equator and the pole), and by determining the pair 
of differential parameters \omeq\ and \dom\ that produces the best fit to the data at 
a given image information content.  
Since differential rotation is estimated from subtle departures with respect to pure 
rotational modulation, we only use LSD photospheric profiles for this task, in order to 
minimise all other potential sources of intrinsic variability.  
When using Stokes $V$ profiles, we find a clear minimum in the \chisqr\ map, located 
at $\omeq=0.974\pm0.005$~\rpd\ and $\dom=0.036\pm0.015$~\rpd\ (see Fig.~\ref{fig:drot}).  
The corresponding rotation periods for the equator and pole are $6.45\pm0.04$~d and 
$6.70\pm0.15$~d respectively, compatible (within $\simeq2\sigma$) with 
the range of reported photometric periods \citep[ranging from 6.35 to 6.60~d][]{Grankin08}.  
This is also compatible (within $\simeq2\sigma$) with the preliminary estimate 
derived from the 2005 Stokes $I$ data set.  
No clear \chisqr\ minimum is observed when using the LSD Stokes $I$ profiles from 
2009 July.  

From our new estimate alone, we cannot firmly exclude that V2129~Oph is rotating rigidly;  
if this were the case, the most likely rotation period (i.e., that minimising \chisqr) would 
be $6.52\pm0.01$~d, close to that assumed here for phasing data (equal to 6.53~d, see Eq.~\ref{eq:eph}).  
Rigid rotation is however unlikely given the observed variability of the photometric period.  
At 99\% confidence level, we can ascertain that the photospheric shear of V2129~Oph is 
Sun-like in sign (i.e.\ with a fast equator and slow pole) and smaller than $\simeq0.07$~\rpd\ 
(about that of the Sun).  The corresponding time for the equator to lap the pole by one 
complete rotation cycle is found to be $175^{+125}_{-52}$~d, and is larger than 90~d at 99\% 
confidence level.  
Images presented in Fig.~\ref{fig:map} are reconstructed assuming our new estimate of the 
photospheric shear;  images derived assuming rigid rotation (with a period of 6.53~d) are 
however virtually identical to those presented here (as expected from the small differential 
rotation).  

\begin{figure}
\includegraphics[scale=0.35,angle=-90]{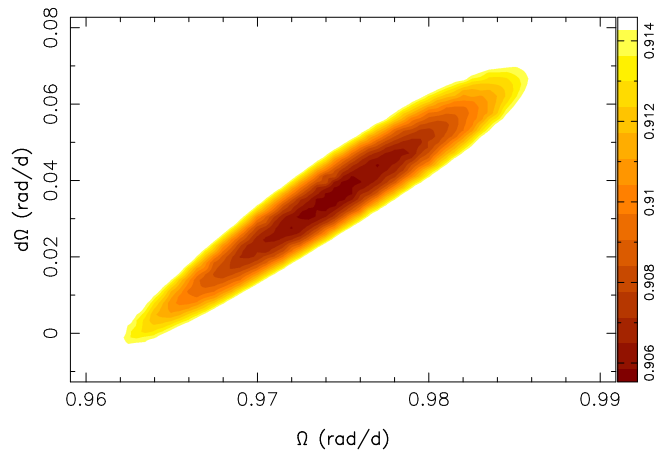}
\caption[]{Variation of \chisqr\ as a function of \omeq\ and \dom, derived from the modelling
of Stokes $V$ LSD profiles at constant information content.  The outer color plot corresponds 
to a \chisqr\ increase of 1\% corresponding to a 99\% confidence interval (for each parameter 
considered separately) given the number of data points adjusted in the process (644). }
\label{fig:drot}
\end{figure}

\section{Summary and discussion}
\label{sec:dis}

This paper presents new results of the MaPP project, focussing on the impact of 
magnetic fields on star formation, during the cTTS phase in particular.  
Here we concentrate on V2129~Oph, the brightest cTTS of the $\rho$~Oph 
star formation cloud, for which we present new spectropolarimetric data 
collected in 2009 July with ESPaDOnS@CFHT, covering 2 complete rotation cycles 
of the protostar and augmented from almost simultaneous photometric data collected 
with SMARTS;  additional multiwavelength data (e.g., X-ray data from Chandra/HETGS, 
visible spectra from HARPS@ESO and SMARTS) were also collected as part of this campaign 
and will be presented in forthcoming companion papers.  
From these time-resolved data and using the last version of our tomographic 
techniques, we reconstruct maps of the large-scale magnetic field, of the 
photospheric brightness and of the accretion-powered \caii\ IRT excess emission at 
the surface of V2129~Oph;  we also include in the paper a complete re-analysis of 
our old spectropolarimetric data collected in 2005 July, to ease comparison 
with our newest results.  

The large-scale field of V2129~Oph is found to be mostly poloidal and axisymmetric, 
and essentially consists of a $\simeq$2.1~kG octupole and a $\simeq$0.9~kG dipole, both 
tilted by about 20\degr\ with respect to the rotation axis;  the main magnetic feature 
reconstructed on the map is a 4~kG high-latitude radial field region.  We also 
find that the large-scale field strongly increased by typically a factor of 2 since 2005 June, with the dipole 
component enhanced by a factor of $\simeq$3 and the octupole component by about 50\%;  
in this process, the octupole to dipole intensity ratio has decreased by a factor of 
$\simeq2$, from about 5 (in 2005) down to $\simeq2.5$ (in 2009).  This change is readily 
visible on the main high-latitude radial field feature, whose average magnetic strength 
(as determined from, e.g., \caii\ IRT or \hei\ lines) has increased by at least a 
factor of 2 between both epochs.  
We finally report that the surface differential rotation by which the detected large-scale 
field is sheared is Sun-like in sign (with a fast equator and slow poles) and weaker 
than solar in strength (by about a factor of 2).  

We consider that the change in the large-scale magnetic topology of V2129~Oph is 
unambiguous evidence that its field is generated by dynamo processes, rather than 
being a fossil remnant of earlier formation stages;  although this conclusion was 
anticipated both on observational and theoretical grounds \citep[e.g.,][]{Donati09}, 
its firm demonstration is nevertheless a significant step forward in our understanding 
of the origin of magnetic fields in low-mass protostars.  
Our result suggests that dynamo fields of cTTSs are variable on a timescale of a few 
years and are thus intrinsically non-stationary;  a regular magnetic monitoring of a small 
sample of low-mass protostars should be able to reveal whether these dynamos are cyclic, 
and if so, provide an estimate of the typical period of their magnetic cycles.  
The recent report that two cTTSs with virtually identical stellar parameters and 
evolutionary states (namely BP~Tau and AA~Tau) show discrepant magnetic 
topologies (with unequal dipole to octupole intensity ratios, respectively equal to 
$\simeq$1 and $>$5 respectively) and different dipole strengths in particular (by 
about a factor of 2), although intriguing at first sight, seems consistent with what we 
observe on V2129~Oph (where we see a comparable change in dipole field strength 
between 2005 and 2009). 

Our data also indicate that areas of enhanced accretion-powered \caii\ and \hei\ emission 
are close to the main magnetic region of V2129~Oph, 
as in all cTTSs monitored with MaPP up to now;  it confirms in particular that this 
strong field region (and its immediate vicinity) is where accretion funnels, 
linking the protostar to the inner edge 
of its accretion disc, are anchored.  We further demonstrate that in V2129~Oph, these 
strong magnetic regions are coincident, at photospheric level, with large dark 
spots featuring a high brightness contrast with the neighbouring photosphere and 
generating most of the observed photometric variations.  We also find that this 
dark spot is increasing in size and contrast between 2005 and 2009, i.e., when the 
field in main magnetic region strengthens, in line with theoretical 
expectations that stronger fields are more efficient at inhibiting local convection.  
We note that the observed photometric modulation of V2129~Oph cannot be attributed to 
partial occultations of the protostar by the accretion warp at the inner disc edge like 
those seen on AA~Tau \citep[e.g.,][]{Bouvier07};  these occultations are indeed generating 
photometric eclipses that are deep (usually larger than 1~mag), color specific \citep{Bouvier99} 
and lagging phases of magnetic maxima \citep{Donati10b}, at variance with what we report here 
for V2129~Oph.  It confirms at the same time that the rotation axis of V2129~Oph is likely 
not inclined by more that 60\degr\ with respect to the line of sight (see Sec.~\ref{sec:v2129}).  

Using the \hei\ $D_3$, \caii\ IRT, H$\beta$ and H$\alpha$ emission fluxes along with the empirical 
correlations of \citet{Fang09}, we derive that the average logarithmic accretion rate of V2129~Oph 
(in \mspy) is $-9.2\pm0.3$, peaking at $-9.0$ in average at magnetic maximum (i.e., phase 0.55).  
It suggests that the mass accretion rate of V2129~Oph is stable\footnote{More data collected over 
a longer time span are however needed to confirm that we did not catch V2129~Oph in an unusually 
quiet state of accretion at both epochs. } to within a factor of $\simeq$2 on 
both short and long timescales (i.e., weeks to years).  Given the intensity of the dipole component 
reconstructed for V2129~Oph (equal to $B_\star\simeq450$~G and $\simeq150$~G at the equator in 2009 
and 2005 respectively, see Sec.~\ref{sec:mod}) and using a logarithmic mass accretion rate of $-9.0$, 
we obtain that the radius \rmag\ at which the 
inner disc of V2129~Oph is truncated is $\rmag\simeq7.2$~\rstar\ and $\simeq3.9$~\rstar\ for 2009 
and 2005 respectively\footnote{Note that our previous estimate of \rmag\ in 2005 June \citep{Donati07}, 
derived from a different expression of \rmag\ and a different mass accretion rate, should be considered 
as obsolete and superseded by the present one.}, or equivalently that $\rmag/\rcor\simeq0.93$ and 
$\simeq0.50$ at the same two epochs \citep[using the theoretical estimates of][]{Bessolaz08}.  

This shows that, in 2009 July, V2129~Oph has a magnetosphere extending out to $\rmag\simeq\rcor$, 
where the dipole component of the field vastly dominates that of the octupole (by a factor of about 
25 in field strength) even though the latter dominates the former at the stellar surface (by about 
a factor of 2.5 and 1.9 at the poles and equator respectively);  in this situation, accretion is 
expected to occur naturally towards the high-latitude magnetic poles only \citep[and not towards 
low-latitudes, as in a pure octupole case, e.g.,][]{Long09}.  Our observations confirm that this is 
indeed the case, with mass accretion at the surface of the star concentrating at a magnetic colatitude 
smaller than 10\degr\ (see Fig.~\ref{fig:map}).  
In 2005 June, we find that the magnetosphere only extends to $\rmag\simeq3.9$~\rstar\ while at the 
same time the dipole field is half as strong as in 2009 with respect to the octupole component 
(the octupole dominating the dipole by a factor of 5 at the poles and 3.8 at the equator, at the 
surface of the star);  we nevertheless find that the dipole still dominates the octupole at \rmag, 
but only by a factor of about 4 in field strength.  We speculate that this is enough for surface 
accretion to concentrate mostly at magnetic poles only (rather than at low latitudes) as suggested 
again by observations.  

Our results are in relatively good agreement with those of recent very detailed 3D MHD simulations 
carried out in the specific case of V2129~Oph \citep{Romanova10}.  In particular, estimates of the 
relative extent of \rmag\ with respect to \rcor\ and its dependence with dipole strength and accretion 
rate are roughly compatible.  The grossly circular shape and polar location of accretion spots found 
in our maps is also similar to theoretical predictions when $\rmag/\rcor\gtrsim0.7$ and differ in 
particular from the predicted crescent-shaped accretion regions in purely dipolar magnetospheres 
\citep[e.g.,][]{Romanova03};  the accretion arcs/rings that simulations predict at lower latitudes 
when $\rmag/\rcor\simeq0.5$ (i.e., as in 2005) are however not observed, for a yet unclear reason.  

Balmer profiles further document the different magnetospheric accretion configurations of V2129~Oph 
in 2009 July and 2005 June.  The standard deviation profiles clearly show that the red-shifted 
absorption in 2009 July occurs at significantly smaller velocities than in 2005 June (see 
Fig.~\ref{fig:bms}).  The origin of this change is not clear yet;  multiple observations 
are needed to investigate this issue in more details.  
In addition, we observe that this red-shifted Balmer absorption is trailing magnetic maximum in 
2009 July (by about 0.15 cycle in average, see Sec.~\ref{sec:var}), while the opposite is observed 
in 2005 June.  We suggest it indicates that part of the accretion funnel is anchored beyond \rcor\ 
in 2009 July (as expected from $\rmag\simeq\rcor$) with the corresponding flux tubes lagging the 
main funnel as they are dragged behind by the slow Keplerian flow;  since $\rmag\ll\rcor$ in 2005 June, 
we may logically expect an opposite situation (with the base of accretion funnels being dragged 
ahead of the main magnetic region by the fast Keplerian flow), in rough qualitative agreement with 
observations.  
Finally, blue-shifted Balmer emission is detected at velocities of $\simeq-170$~\kms\ in 2009 
July while none was seen in 2005 June, suggesting that V2129~Oph is now triggering a significant 
outflow;  we propose that this outflow comes from the inner disc, tracing the (small) fraction of 
the accreting disc material expelled outwards, as expected when $\rmag\simeq\rcor$.  

We also report that the high-latitude accretion-powered \caii\ IRT excess emission spot that we detect 
at the surface of V2129~Oph (at phase $0.6-0.7$) is slightly trailing the main magnetic region as well as the 
accretion-powered \hei\ emission spot (both centred at phase 0.5) by about 0.1 cycle in 2009 July (see 
Sec.~\ref{sec:var}), while all three regions were more or less overlapping in 2005 June.  
The origin of this small (but apparently real) spatial shift is not clear yet.  
We suspect that this phenomenon relates to the trailing accretion veils 
reported above, even though red-shifted Balmer absorption obviously traces a physically different 
magnetospheric plasma than \caii\ IRT excess emission;  we propose for instance that it reflects the 
complex temperature structure of the postshock region that results from the sheared accretion funnels.  
More data are obviously needed to investigate this issue in a quantitative way.  
 
As a summarising conclusion, we propose a schematic description of how magnetic fields are likely 
to affect the evolution of cTTSs, and in particular their angular momentum.  
Our new spectropolarimetric data set on V2129~Oph shows that magnetic fields of cTTSs exhibit a 
large-scale structure, and in particular a dipolar component, that varies on a timescale of a few 
years, as a likely result of a non-stationary dynamo.  In this context, magnetospheric 
accretion, essentially controlled by the lowest orders of the large-scale field of the protostar, 
is expected to reflect more or less directly changes in the magnetic topology.  
When the dipole component of the field is strong enough to ensure that $\rmag\gtrsim\rcor$, the 
protostar enters a propeller-like regime and is actively spun-down;  when the dipole field gets 
weaker, \rmag\ shrinks back within \rcor\ and the star spins up again.  Since the typical 
fluctuation timescale of dynamo fields (of order of years to decades) is much shorter than that 
of star-disc angular momentum transfer (by at least 4 orders of magnitude), the protostar ends 
up rotating at an average rate;  this rate more or less reflects the mean dipole strength 
(with larger fields implying slower rotation) and mean mass accretion rate (with stronger 
accretion implying faster rotation) over the averaging timescale.  As a result, cTTSs with 
masses in the range $0.5-1.3$~\msun\ rotate slowly, their dynamo fields being strongly 
dipolar and axisymmetric like those of mid-M dwarfs \citep{Morin08b};  cTTSs with either lower 
or larger masses rotate more rapidly, being apparently less efficient at producing strong 
dipole fields \citep[e.g.,][]{Donati10}.  AA~Tau and BP~Tau obviously belong to the first category, 
both hosting a dipole component stronger than 1~kG \citep{Donati08, Donati10b}.  
With a radiative core growing in size rapidly, V2129~Oph is apparently no longer successful at 
building up a strong dipole field (or no more than for a small amount of time) and can likely no 
longer counteract its spin-up.  

Upon completion, MaPP should be able to assess (on a small statistical sample) how well our 
proposed picture matches cTTSs;  in particular, repeated and regular visits on a limited, 
well-selected subgroup, like that presented here for V2129~Oph, will be essential to address 
the important issue of non-stationary dynamos in low-mass protostars, to evaluate the constant 
and fluctuating parts of their dipolar fields, and to quantify the associated impact on their 
angular momentum evolution.

\section*{Acknowledgements}
This paper is based on 
observations obtained at the Canada-France-Hawaii Telescope (CFHT), operated by the National 
Research Council of Canada, the Institut National des Sciences de l'Univers of the Centre 
National de la Recherche Scientifique of France and the University of Hawaii. 
The ``Magnetic Protostars and Planets'' (MaPP) project is supported by the 
funding agencies of CFHT and TBL (through the allocation of telescope time) 
and by CNRS/INSU in particular, as well as by the French ``Agence Nationale 
pour la Recherche'' (ANR).
We thank an anonymous referee for suggesting various improvements to the paper, and 
the CFHT/QSO and TBL staff for their efficiency at collecting data.  
SGG acknowledges support by the Science and Technology
Facilities Council [grant number ST/G006261/1].  

\bibliography{v2129}
\bibliographystyle{mn2e}
\end{document}